\documentclass[twocolumn,showpacs,preprintnumbers,amsmath,amssymb]{revtex4}

\usepackage{graphicx}
\usepackage{dcolumn}
\usepackage{bm}

\def\bea{\begin{eqnarray}}
\def\eea{\end{eqnarray}}

\begin{document} 

\preprint{Version 2.2}


\title{Evolution of minimum-bias parton fragmentation in nuclear collisions}

\author{Thomas A. Trainor}
\address{CENPA 354290, University of Washington, Seattle, WA 98195}


\date{\today}

\begin{abstract}
Minimum-bias fragment distributions (FDs) are calculated by folding a power-law parton energy spectrum with parametrized fragmentation functions (FFs) derived from $e^+$-$e^-$ and p-\=p collisions. Changes in FFs due to parton ``energy loss'' or ``medium modification'' are modeled by altering FF parametrizations consistent with rescaling QCD splitting functions. The common parton spectrum is constrained by comparison with a p-p $p_t$ spectrum hard component. In-vacuum and in-medium FDs are compared with spectrum hard components from 200 GeV Au-Au collisions for several centralities. The reference for all nuclear collisions is the FD derived from in-vacuum $e^+$-$e^-$ FFs. The hard component for p-p and peripheral Au-Au collisions is found to be {\em strongly suppressed} for smaller fragment momenta, consistent with the FD derived from in-vacuum p-\=p FFs. At a particular centrality the Au-Au hard component transitions to enhancement at smaller momenta and suppression at larger momenta, consistent with FDs derived from  in-medium $e^+$-$e^-$ FFs. Fragmentation systematics suggest that QCD color connections change dramatically in more-central A-A collisions. Spectrum systematics are inconsistent with saturation-scale arguments and parton thermalization. 
\end{abstract}

\pacs{12.38.Qk, 13.87.Fh, 25.75.Ag, 25.75.Bh, 25.75.Ld, 25.75.Nq}

\maketitle

 \section{Introduction}

RHIC collisions are conventionally described in terms of two major themes: hydrodynamic evolution of a thermalized bulk medium~\cite{finns,heinz,rom,bwfit} and energy loss of energetic partons in that medium {\em via} gluon bremsstrahlung~\cite{highpt}. Medium dynamics and properties and parton specific energy loss relating to ``tomography'' of the medium are the principal analysis goals~\cite{tomo,tomo2}. Analysis methods tend to favor those goals: Methods directed toward a bulk medium tend to suppress low-$p_t$ features of parton fragmentation, and methods applied to high-$p_t$ jet analysis also tend to suppress structure at smaller $p_t$~\cite{2comp}. 

Recent physical-model-independent studies of spectrum and correlation structure have revealed interesting new aspects of RHIC collisions. Analysis of number and $p_t$ angular correlations led to unanticipated structure in the final state, subsequently identified with parton fragmentation in the form of {\em minijets}~\cite{axialci,hijscale,ptscale,ptedep,daugherity}. Two-component analysis of p-p and Au-Au spectra revealed a corresponding {\em hard component}, a minimum-bias fragment distribution associated with minijets which suggested that jet phenomena extend down to 0.1 GeV/c~\cite{ppprd,2comp}.

In this analysis new aspects of spectra and correlations in p-p and Au-Au collisions are combined with complete representations of fragmentation functions from $e^+$-$e^-$~\cite{ffprd} and p-\=p~\cite{cdf1,cdf2} collisions to reveal the systematic evolution of parton fragmentation with Au-Au centrality. Accurately parametrized FFs are combined with a power-law parton spectrum to produce fragment distributions which can be compared quantitatively with hard components derived from $p_t$ spectra in nuclear collisions. The observed FD evolution reveals surprising new features of parton fragmentation in p-p and A-A collisions.

\section{Minijets}

Minijets dominate the transverse dynamics of nuclear collisions above $\sqrt{s_{NN}} \sim$ 15 GeV. They have an experimental and theoretical history of more than twenty years. 
The term ``minijets'' can be applied collectively to all hadron fragments from the minimum-bias scattered-parton spectrum  averaged over a given A-A or N-N event ensemble. Because the parton $p_t$ spectrum is rapidly varying ($\sim 1/p_t^{7}$), the minimum-bias spectrum is nearly monoenergetic, peaked at an effective termination or cutoff energy near 3 GeV~\cite{ua1,kll,sarc}. The term ``minijets'' then refers {\em experimentally} to jets localized near the cutoff energy.

Minijets manifest as both minimum-bias {jet correlations}~\cite{jeffpp} and as the {hard component} of the two-component spectrum model~\cite{ppprd,2comp}. They  provide {unbiased} access to fragment distribution structure down to a small cutoff energy for scattered partons (those fragmenting to charged hadrons) and to the smallest detectable fragment momenta ($\sim 0.1$ GeV/c). Because they are large enough to observe accurately but small enough to respond fully to any QCD medium, minijets serve as Brownian probes of QCD in nuclear collisions ~\cite{newflow}.

The minijet concept emerged experimentally at the  SP\=PS from a UA1 analysis of $E_t$ structure down to small integrated $E_t$~\cite{ua1}. The analysis determined that $E_t$ ``clusters'' (minijets) are distributed according to the expected pQCD power-law parton spectrum down to 5 GeV. Azimuth correlations {\em between} clusters exhibited a peak at $\pi$ radians expected for back-to-back scattering of initial-state partons. The 5 GeV $E_t$ cutoff was later related to a 3-4 GeV parton energy equivalent~\cite{sarc}.

Corresponding minijet structure was observed in two-particle correlations from 200 GeV p-p collisions~\cite{jeffpp}. Angular correlations with no ``jet'' $p_t$ conditions exhibit just the structure expected from pQCD jets: a narrow {\em intrajet} same-side peak at the angular origin (parton fragmentation) with most-probable $p_t \sim 1$ GeV/c and an {\em interjet} away-side ridge at $\pi$ radians (back-to-back parton scattering). As noted by UA1, there is no dividing line between conventional high-$p_t$ ``jets'' and ``minijets.''

Several theoretical treatments have identified experimental minijets with parton scattering and fragmentation. Minijet production was calculated perturbatively for anticipated RHIC U-U collisions based on the UA1 minijets: ``The observed [minijet] rate is in agreement with [p]QCD and is quite large''~\cite{kll}. ``Semihard parton interactions [as in the UA1 minijet analysis] appear to play an important role in high-energy hadronic scattering [$\sqrt{s_{hh}} \gg 10$ GeV]''~\cite{durand}.  A ``...theoretical cutoff of $p_t^{min} \sim 3$ GeV seems to describe the observed total minijet cross section with $E_T^{jet} (E_T^{raw}) \geq 5$ GeV'' and produces a minijet total cross section in agreement with UA1 data~\cite{sarc}. ``There is an increasing amount of evidence that the perturbative domain of QCD extends down to [parton] momenta of the order of 1 GeV/c''~\cite{blaizot,lowpqcd}.

Minijet thermalization is of central interest as the basis for QGP formation and hydrodynamic flows. ``Minijets...will be reprocessed by the system and not emerge from it''~\cite{kll}. In~\cite{nayak} the thermalization time is estimated as 4-5 fm/c with $T \sim 200$ GeV. One basis for claims of thermalization is the assumption that partons (gluons) propagate in a gas of bare gluons, which can be strongly questioned given recent minijet-related results at RHIC (e.g., the present analysis and~\cite{axialci,jeffpp,ptscale,ptedep,daugherity}).

The HIJING Monte Carlo was developed specifically to study the role of minijets in p-p and A-A collisions:  ``We emphasize the effects due to multiple mini-jet production at collider energies''~\cite{hijingpp,minijet}. HIJING p-p correlations quantitatively match minijet correlations (same-side amplitude, widths, away-side ridge) measured in p-p collisions~\cite{jeffpp}. HIJING predictions with ``jet quenching'' disabled are consistent with a {\em Glauber linear superposition} reference for A-A collisions~\cite{ptscale,daugherity}. 

\section{Analysis method}

In this analysis minijets manifested as $p_t$-spectrum hard components are the object of study. Experimental hard components are modeled with fragment distributions calculated by folding parton spectra with various fragmentation-function ensembles. From the comparisons parton spectrum parameters and modifications to fragmentation functions in more-central Au-Au collisions are inferred.

A study of the charge-multiplicity $n_{ch}$ dependence of p-p $p_t$ spectra revealed two components with fixed functional forms independent of $n_{ch}$, denoted soft and hard components and later interpreted in terms of longitudinal projectile fragmentation (soft) and transverse scattered-parton fragmentation (hard)~\cite{ppprd}. Separation into two components was based on a Taylor expansion of spectrum structure on event $n_{ch}$ with no physical model imposed. The hard-component fragment distribution or FD was identified with minijet angular correlations, having all the characteristics of jet correlations but with no jet-specific $p_t$ cuts imposed (i.e., minimum-bias jets)~\cite{jeffpp}.

Analysis of $e^+$-$e^-$ (e-e) fragmentation functions (FFs) from LEP and HERA led to a complete characterization of FFs for all parton energy scales in terms of beta distributions on normalized rapidity $u$. FFs were represented to their statistical limits for all fragment momenta (in contrast to the limitations of conventional pQCD parametrizations such as the MLLA)~\cite{ffprd}. The most important achievement was accurate representation of FFs near the smallest fragment and parton momenta.  Comparison with p-\=p (p-p) FFs indicated substantial systematic e-e vs p-p differences for smaller fragment momenta not revealed by conventional data plots on momentum fraction $x_p = p / p_{jet}$ or its logarithmic equivalent $\xi_p = \ln(1/x_p)$. 

In this analysis e-e and p-p FFs are folded with a parton spectrum model to produce fragment distributions to be compared with measured spectrum hard components. For example, a measured FF ensemble from p-\=p collisions is folded with the spectrum model to produce a calculated FD. Comparison of the FD with a measured p-p spectrum hard component determines the spectrum-model cutoff energy and QCD power-law exponent. The parton spectrum agrees quantitatively with pQCD predictions and is comparable with a UA1 measurement of the differential jet cross section based on $E_t$ clusters. 

The question then arises which FFs should be used to calculate FDs for various nuclear collision conditions. In the initial part of the analysis it is assumed that the underlying parton spectrum remains unchanged for a given final-state hadron acceptance (e.g., $p_t \geq 0.15$ GeV/c, $|\eta| < 1$, $2\pi$ azimuth). Manifestations of ``jet quenching'' or parton ``energy loss'' are modeled via FF modifications accessible for the first time over the full $p_t$ acceptance. A scheme for FF medium modification in~\cite{b-w} is found to be particularly relevant to data. 

Spectrum ratio measures and direct comparisons of hard components with calculated FDs reveal the relation between systematic FF modifications and Au-Au centrality. Ironically, the most significant fragmentation changes occur below $p_t \sim 2$ GeV/c where most fragments appear but where comparisons with pQCD are typically deferred in favor of hydrodynamic descriptions. In particular, a direct correspondence has emerged between dramatic fragmentation modifications noted in this analysis and the recently-observed {\em sharp transition} in minijet correlation systematics at a specific Au-Au centrality~\cite{daugherity}.

\section{p-p two-component spectra}

The two-component model of p-p spectra~\cite{ppprd} was the starting point for the differential fragmentation analysis described in this paper. The two-component model emerged from a Taylor-series expansion of spectra on observed charge multiplicity $\hat n_{ch}$ in one unit of $\eta$ ($\sim d\hat n_{ch}/d\eta$) and was not motivated by a particular physical model. The spectrum components (Taylor series coefficients) were subsequently interpreted physically in the context of correlation analysis and by analogy with parton fragmentation systematics at larger energy scales.

\subsection{Two-component spectrum model}

The two-component model applies to two-particle correlations and to their 1D projections, the $p_t$ or $y_t$ spectra. The two-component spectrum model for p-p collisions sorted according to event-multiplicity index $\hat n_{ch}$ is
\bea \label{2compalg}
 \frac{1}{n_s(\hat n_{ch})}\frac{1}{y_t}\, \frac{dn_{ch}(\hat n_{ch})}{dy_t } =  S_0(y_t)  +  \frac{n_h(\hat n_{ch})}{n_s(\hat n_{ch})}\,  H_{0}(y_t),
\eea
where $n_x$ is integrated over one unit of $\eta$ (i.e., $n_x / 2\pi \sim d^2n_x/d\eta d\phi$), soft component $S_0(y_t)$ is the Taylor series ``constant,'' and hard component $H_0(y_t)$ is the coefficient of the term linear in $\hat n_{ch}$, both normalized to unit integral. For comparisons with A-A spectrum data we define $S_{pp} =(1/y_t)\, dn_s/dy_t$ with reference model $n_s\, S_0$ and similarly for $H_{pp} \leftrightarrow n_h\, H_0$. The two-term Taylor series exhausts all significant p-p spectrum structure. 

\begin{figure}[h]
  \includegraphics[width=1.65in,height=1.65in]{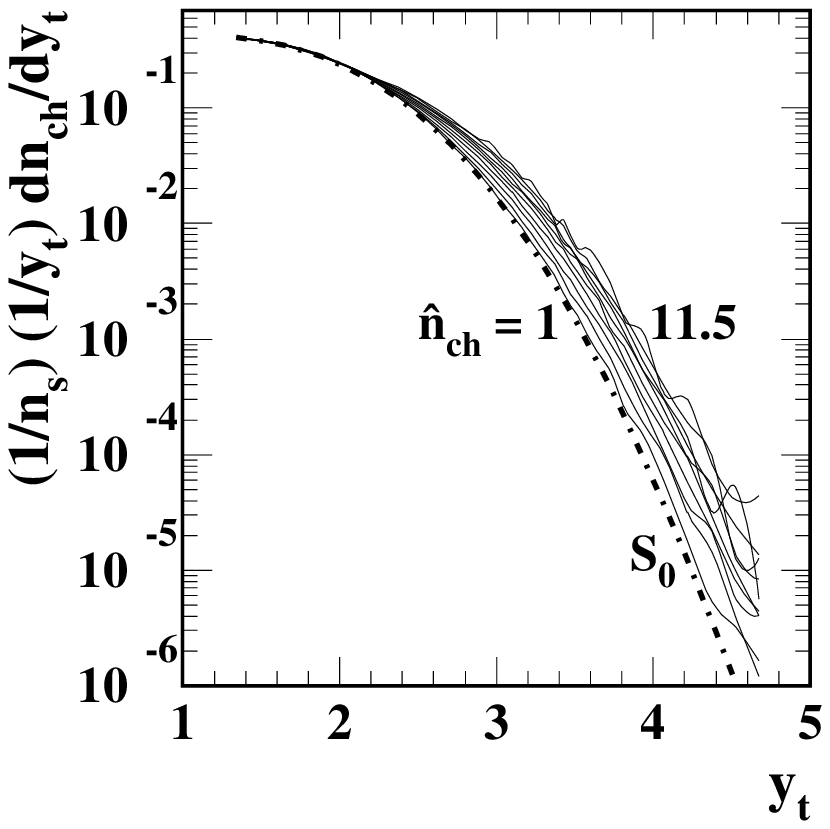}
  \includegraphics[width=1.65in,height=1.65in]{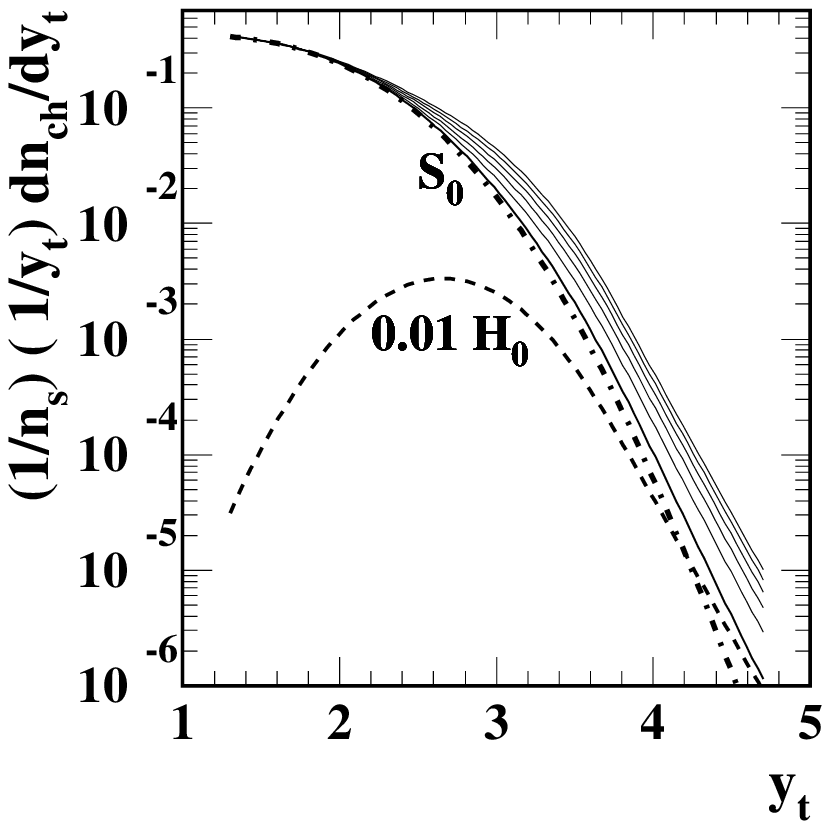}
\caption{\label{ppspec}
 Left: Spectra for ten multiplicity classes [1,11.5] of 200 GeV NSD p-p collisions~\cite{ppprd}. The dash-dotted curve is the spectrum soft component $S_0(y_t)$ defined as the limiting case for $\hat n_{ch} \rightarrow 0$.
Right: The two-component (soft+hard) model of p-p spectra. Hard component $H_0(y_t)$ is a Gaussian with QCD power-law tail~\cite{2comp}.
} 
 \end{figure}

Fig.~\ref{ppspec} (left panel) shows spectra for ten multiplicity classes from 200 GeV  non-single diffractive (NSD) p-p collisions~\cite{ppprd}. The asymptotic limit for $\hat n_{ch} \rightarrow 0$ (dash-dotted curve) is $S_0$. The spectra are normalized by the soft-component multiplicity $n_s = n_{ch} / (1 + \alpha\, \hat n_{ch})$, where $\alpha \sim 0.01$ and $\hat n_{ch} \sim n_{ch} / 2$ is the observed $n_{ch}$ resulting from incomplete $p_t$ acceptance and tracking inefficiencies.

Fig.~\ref{ppspec} (right panel) shows the two-component algebraic model Eq.~(\ref{2compalg}) with unit-normal model functions $S_0$ and $H_0$ defined in~\cite{ppprd,2comp}. The hard-component spectrum contribution $n_h  / n_s$ scales as $\alpha\, \hat n_{ch}$. Factor $\alpha = 0.01$ is the average value for most $\hat n_{ch}$ classes. The factor drops to 0.0055 for  $\hat n_{ch} = 1$. The spectrum data in the left panel are described to the statistical limits.

\subsection{p-p spectrum hard component}

\begin{figure}[h]
\includegraphics[width=.47\textwidth,height=1.75in]{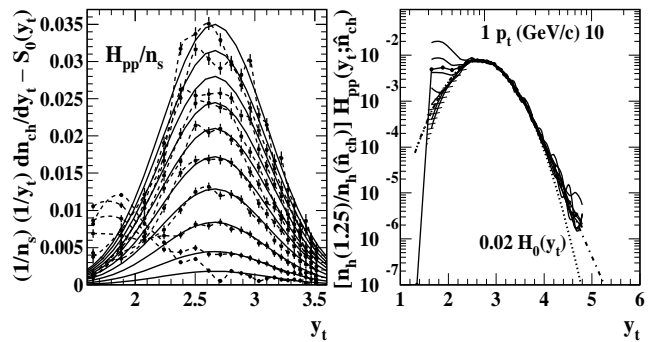}
\caption{\label{pphard} 
Left: Spectrum hard components $H_{pp}(y_t,\hat n_{ch})$ (solid points) for ten multiplicity classes of 200 GeV p-p collisions~\cite{ppprd}. A common Gaussian model function describes the data well except for parts of the lowest multiplicities. Right: $H_{pp}$ data from the left panel normalized to NSD p-p collisions by factor $n_h(1.25)/n_h(\hat n_{ch})$ demonstrating the common form. A Gaussian (dotted curve) and Gaussian with power-law tail (dash-dotted curve) are compared to the data.
} 
\end{figure}

Figure~\ref{pphard} (left panel) shows p-p hard components in the form $ H_{pp} / n_s$ for ten multiplicity classes obtained by subtracting fixed soft component $S_0$ from the ten NSD p-p spectra normalized by soft multiplicity $n_s$. The hard-component shape is independent of multiplicity and described approximately by a Gaussian. The amplitude is approximately proportional to $\hat n_{ch}$. From the figure $H_0$ coefficient $n_h / n_s = \alpha\,  \hat n_{ch}$ is inferred, with $\alpha \sim 0.01$~\cite{ppprd}.

Figure~\ref{pphard} (right panel) shows ten hard components $H_{pp}$ from the left panel scaled by factors $n_h(1.25) / n_h(\hat n_{ch})$ to reveal a common shape representing the mean hard component for NSD p-p collisions. The dash-dotted curve is $0.02\, H_0$ [$0.02 \sim (\alpha = 0.007)\, (\hat n_{ch}=1.25)\, (n_{s} = 2.5)$~\cite{ppprd}], with $H_0$ defined as a Gaussian plus exponential tail on transverse rapidity $y_t$ ($n_h\, H_0$ is relabeled below as reference $H_{GG}$). The exponential tail represents the QCD power law $\propto p_t^{n_{QCD}} \rightarrow \exp[(n_{QCD}-2)\,y_t]$, where the $-2$ results from the $p_t \rightarrow y_t$ Jacobian~\cite{2comp}. The dotted curve, a Gaussian with no QCD tail~\cite{ppprd}, is inconsistent with data at larger $y_t$. The spectrum hard component is interpreted as a {\em minimum-bias} fragment distribution dominated by ``minijets''---jets from those partons (gluons) with at least the minimum energy required to produce charge-neutral combinations of charged hadrons.

\section{Fragmentation functions}

e-e and p-p FFs for inclusive fragments and inclusive partons have been parametrized accurately over the full $(y,y_{max}) \leftrightarrow (x_p,Q^2)$ region relevant to nuclear collisions. The parametrizations permit comprehensive tests of pQCD in relation to nuclear collision data.

\subsection{Accurate FF parametrizations}

Fragmentation functions provide direct experimental access to the parton-hadron interface of QCD. At RHIC energies nuclear collisions are dominated by parton scattering and fragmentation. For full understanding of collision dynamics FFs should be described over the entire fragment distribution and over all parton energies relevant to nuclear collisions. 

pQCD theory emphasizes the parton splitting cascade and DGLAP evolution of FFs. The 10\% most-energetic fragments are well-described (e.g.,~\cite{kkp}). FFs are conventionally represented by semilog plots of $D(x_p,Q^2)$ on momentum fraction $x_p = p / p_\text{parton}$, which compress and obscure details at small $x_p$ (including most of the fragment distribution) or on $\xi_p = \ln(1/x_p)$ for which the lower limits of FFs on $p$ are not well-defined. In contrast, rapidity $y = \ln[(E + p)/m_\pi]$ is well-defined as fragment $p \rightarrow 0$ ($y \rightarrow p/m_\pi$), which is essential for study of FFs in nuclear collisions. The small-$x_p$ (small-$y$) region dominates nuclear collisions, both by driving collision dynamics and by providing diagnostic evidence in the final state.

Two goals should be distinguished: 1) phenomenological descriptions of FFs which can provide simple and accurate representations of data over the large kinematic intervals required for comprehensive study of nuclear collisions; 2) theoretical descriptions of FFs via pQCD, which may be limited by the current state of theory, 

\subsection{$\bf e^+$-$\bf e^-$ fragmentation functions}

$e^+$-$e^-$ light-quark (uds) and gluon fragmentation functions are well-described above energy scale (dijet energy) $Q \sim 10$ GeV by a two-parameter beta distribution $\beta(u;p,q)$ on scaled rapidity $u$~\cite{ffprd}. Parameters $(p,q)$ vary slowly and linearly with $Q$ above 10 GeV and can be extrapolated reasonably well down to $Q \sim 4$ GeV based on dijet multiplicity data. Most of the FF ``scaling violations'' described by the DGLAP equations result from self-similar variations of FFs with energy scale, which can be absorbed into dijet multiplicity $2\, n(Q)$ and scaled rapidity $u \equiv (y - y_{min})/(y_{max} - y_{min})$,  with $y_{max} \equiv \ln(Q / m_\pi)$ and fixed $y_{min}$ inferred from the systematics of measured $e^+$-$e^-$ FFs. Dijet multiplicities are determined by $\beta(u;p,q)$ according to an energy sum rule.

\begin{figure}[h]
\includegraphics[width=1.65in,height=1.75in]{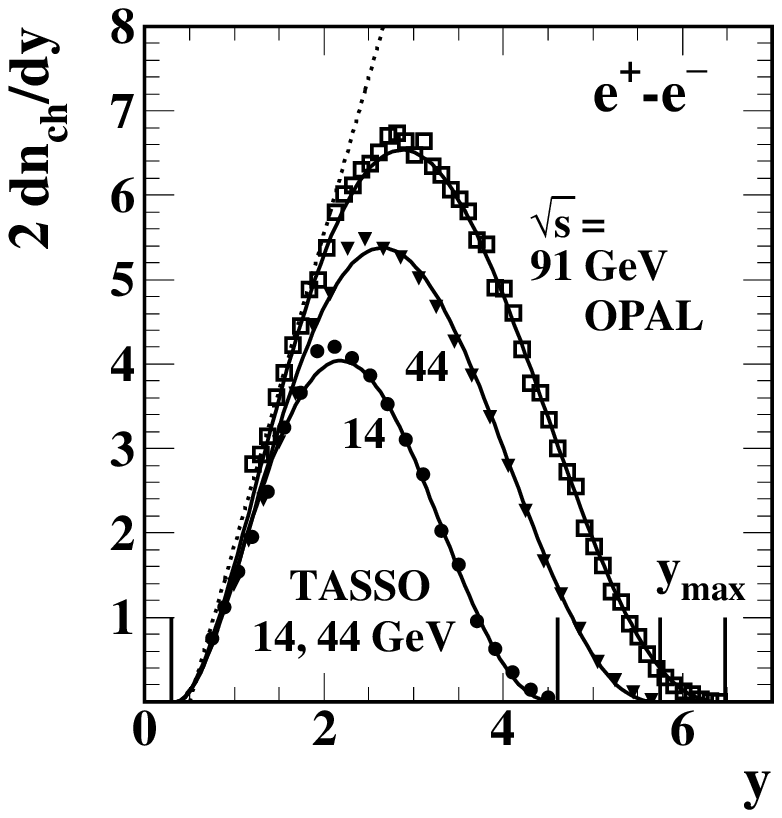}
\includegraphics[width=1.65in,height=1.75in]{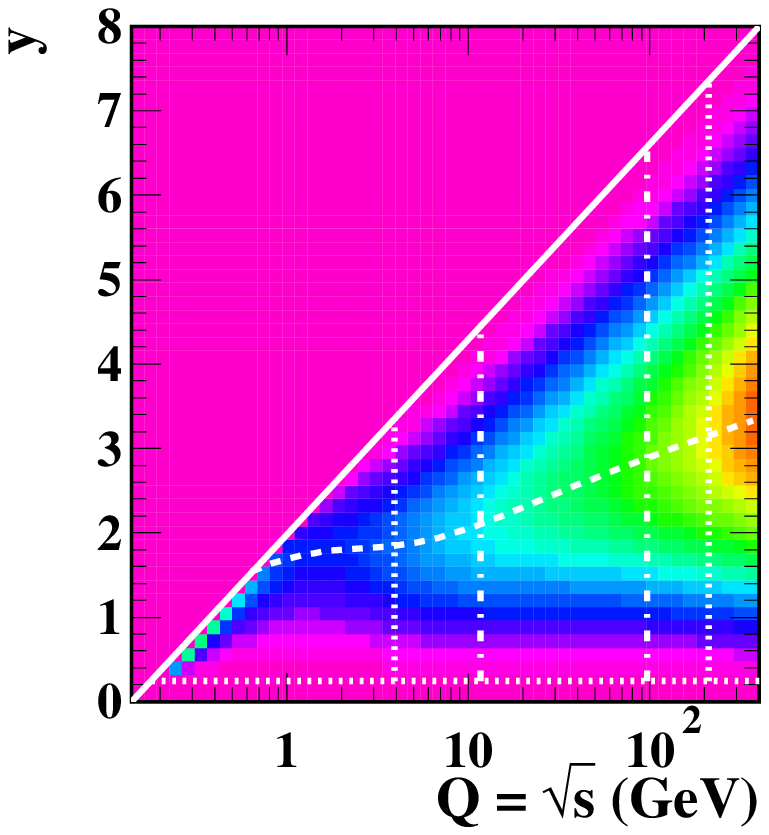}
\caption{\label{eeffs} 
Left: Measured $e^+$-$e^-$ fragmentation functions (symbols) for three energy scales (dijet energies)~\cite{opal,tasso}. Curves through data are from a universal parametrization based on the beta distribution~\cite{ffprd}. Right: (Color online) Surface plot of the universal FF parametrization showing the locus of modes (dashed curve).
} 
\end{figure}

Fig.~\ref{eeffs} (left panel) shows measured FFs for three energy scales from HERA/LEP~\cite{opal,tasso}. The 2 in the axis label indicates that these are dijet $n_{ch}$ densities. The vertical lines at right denote $y_{max}$ values. The curves are determined by the $(p,q)$ parametrization with $y_{min} \sim 0.35$ ($p_t \sim 0.05$ GeV/c, left vertical line) and describe data to their error limits over the entire fragment momentum range. 

Fig.~\ref{eeffs} (right panel) shows the FF ensemble (inclusive light quarks fragment to inclusive hadrons) vs energy scale $Q$ as a surface plot. The dashed curve is the {\em locus of modes}---the maximum points of the FFs. Between the dash-dotted lines the system is determined by fiducial FF data (i.e., exceptional quality and range~\cite{ffprd})  Between the dash-dotted and dotted lines the parametrization is constrained only by dijet multiplicities. 

Systematic trends can be extrapolated to the left of the left dotted line. At the end of a splitting cascade partons evolve to color-singlet hadrons 1-to-1 with a single hadron in the parton ``fragmentation function'' and $y_{fragment} = y_{parton}$ (diagonal line). Approach of the locus of modes to the diagonal therefore leads to local parton-hadron duality (LPHD)~\cite{lphd}. At lower energy scales the QCD density of states is small and hadron resonances dominate. The QCD splitting cascade transitions to a resonance cascade terminating in detected hadrons.


\begin{figure}[h]
\includegraphics[width=3.3in,height=1.75in]{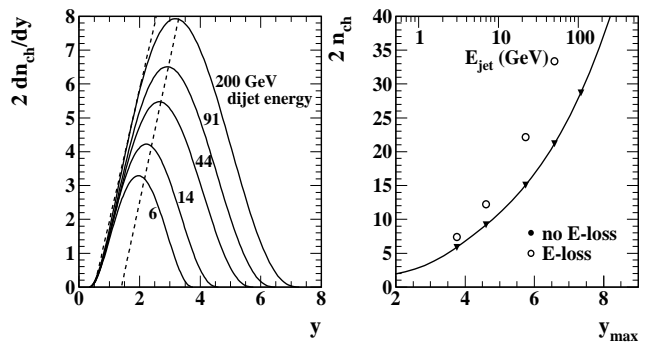}
\caption{\label{eeffs2} 
Left: parametrized $e^+$-$e^-$ fragmentation functions for five dijet energies. Right: Integrated dijet multiplicities for in-vacuum FFs in the left panel (solid points) and for FFs modified according to a model of ``energy loss'' or medium modification~\cite{b-w} in central Au-Au collisions (open circles).
} 
\end{figure}

Figure~\ref{eeffs2} (left panel) shows parametrized beta FFs for five energy scales. The 6-GeV scale is relevant to the minijet spectrum and fragment distributions from this analysis. Such curves provide a complete description of e-e FFs at energy scales relevant to nuclear collisions.

Figure~\ref{eeffs2} (right panel) shows light-quark dijet multiplicity systematics from  the same beta parametrization. The solid points correspond to the FFs in the left panel. The open circles represent multiplicities from medium modification of those FFs in central Au-Au collisions at 200 GeV, as described in Sec.~\ref{bweloss}. The ``in-medium'' shift of FFs to smaller fragment momenta requires more fragments to satisfy energy conservation. The systematics of quark and gluon jets coincide for energy scales $Q = 2 E_{jet} < 8$ GeV. Quark-gluon differences at larger energy scales are less important for gluon-dominated minimum-bias fragmentation in nuclear collisions.

\subsection{p-\=p fragmentation functions}

\begin{figure}[h]
\includegraphics[width=1.65in,height=1.75in]{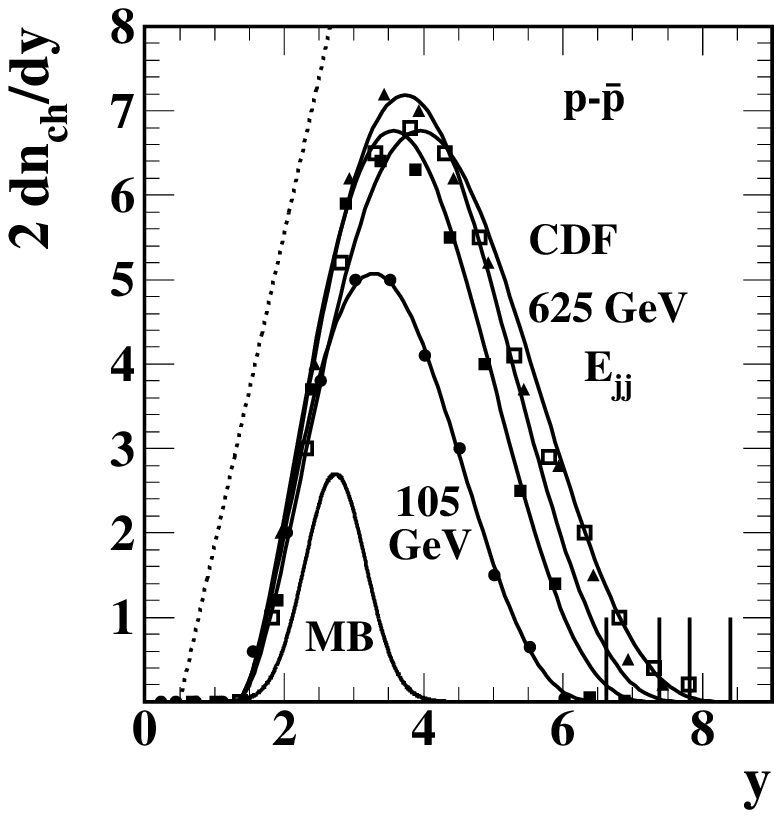}
\includegraphics[width=1.65in,height=1.75in]{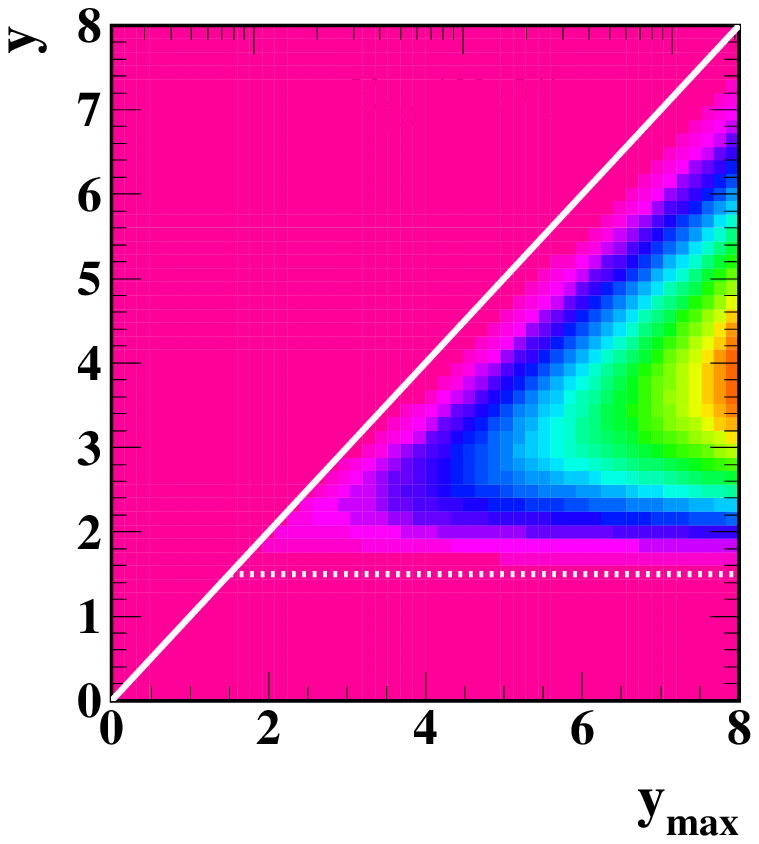}
\caption{\label{ppffs} 
Left: Measured fragmentation functions (samples) from p-\=p collisions at several energies (symbols)~\cite{cdf1}. Solid curves guide the eye. The dotted line represents contrasting $e^+$-$e^-$ FF systematics. Right: (Color online) Surface plot of the universal $e^+$-$e^-$ FF beta parametrization modified with a common cutoff factor to describe p-\=p FFs.
} 
\end{figure}

\begin{figure}[h]
\includegraphics[width=3.3in,height=1.75in]{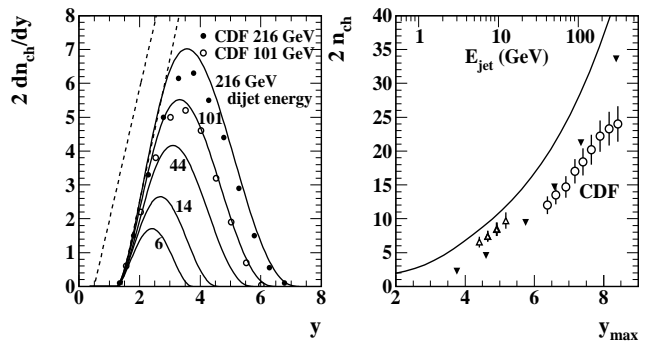}
\caption{\label{ppffs2} 
Left: Parametrized p-\=p fragmentation functions for five dijet energies (solid curves). The FF data (samples) at two energies (symbols)~\cite{cdf2} reveal significant suppression relative to the $e^+$-$e^-$ beta parametrization. Right: Calculated dijet multiplicities (solid points) for in-vacuum p-\=p FFs  (solid curves in left panel) showing significant reductions from the in-vacuum $e^+$-$e^-$ trend (solid curve). Open triangles~\cite{cdfmult} and open circles~\cite{cdf3} are CDF measured p-\=p dijet multiplicities.
} 
\end{figure}

Figure~\ref{ppffs} (left panel) shows FF data from p-\=p collisions at FNAL~\cite{cdf1}. The plotted points are samples from the full data set. Jets are integrated within a cone half-angle of 0.47 radians. The solid curves guide the eye. There is a significant systematic difference between p-p and e-e FFs. The dotted line represents the lower limit for e-e FFs. The systematic gap for all parton energies is apparent---$y_{min}$ for p-p collisions is $\sim 1.5$ (0.3 GeV/c) instead of 0.35 (0.05 GeV/c). The curve labeled MB is hard-component reference $H_{GG}$ from p-p collisions~\cite{ppprd}, comparable to the 6 GeV FF curve in Fig.~\ref{ppffs2} (left panel).

Figure~\ref{ppffs} (right panel) shows a surface plot of the p-p FF ensemble. For a systematic  representation of p-p FFs the e-e FF beta parametrization has been modified by adding a cutoff factor
\bea
g_\text{cut}(y) = \tanh\{ (y - y_0)/\xi_y\}~~~y > y_0,
\eea
with $y_0 \sim \xi_y \sim 1.6$ determined by the CDF FF data~\cite{cdf1}. The modified e-e FFs have not been rescaled to recover the initial parton energy. The cutoff function represents real fragment and energy loss from p-p relative to e-e FFs.

Figure~\ref{ppffs2} (left panel) shows e-e beta FFs for five parton energies~\cite{ffprd} modified by the $g_\text{cut}$ factor. Also plotted are more-recent CDF FF data for dijet energies 101 and 216 GeV~\cite{cdf2} demonstrating the correspondence. The CDF FFs also reveal a systematic amplitude saturation or suppression at larger parton energies compared to LEP systematics, evident also in Fig.~\ref{ppffs} (left panel). In Fig.~\ref{ppffs2} (left panel) the 216 GeV p-p data fall well below the 216 GeV LEP expectation (solid curve). The 101 GeV p-p data fall below the 101 GeV e-e parametrization to a lesser degree, and mainly for smaller fragment rapidities.

Figure~\ref{ppffs2} (right panel) shows multiplicity systematics (solid points) for p-p (i.e., modified e-e) FFs. The solid curve represents unmodified e-e FFs. There is substantial reduction of p-p FF multiplicities due to the cutoff. Also plotted are CDF FF multiplicities from reconstructed jets (open triangles~\cite{cdfmult} and open circles~\cite{cdf3}). 

At 100 GeV dijet energy ($E_{jet} = 50$ GeV) the FF multiplicity in p-p collisions is reduced by $\sim 6$ relative to e-e FFs, with missing-fragment mean $p_t  \sim 0.4$ GeV/c. The corresponding $\sim 2.5$ GeV missing energy represents a small fraction of the total jet energy (possibly within a calorimeter calibration error). But the 30\% $n_{ch}$ reduction could have a major impact on the description of nuclear collisions. At smaller energy scales the fractional multiplicity reduction is much larger (e.g., 70\% for $E_{jet} = 3$ GeV). We return to that important issue in Sec.~\ref{production}.

\section{Fragment distributions}

Whereas a fragmentation function (FF) is conditional on a specific parton energy, a {\em fragment distribution} (FD) is the hadron distribution associated with a minimum-bias parton spectrum---the folding of an FF ensemble with the parton spectrum. The hard component from p-p spectra can be interpreted as an FD~\cite{ppprd}, consistent with p-p correlation systematics~\cite{jeffpp}. This analysis provides further support for the FD interpretation.

NLO ``fragmentation functions''~\cite{nlo} combine a pQCD parton spectrum with a theoretical parametrization of {in-vacuum} e-e FFs, e.g.~as measured at LEP/HERA~\cite{opal,tasso}. Two questions arise: 1) Is the theory description of FFs adequate over the entire fragment momentum range and parton energy range relevant to nuclear collisions? 2) Are e-e FFs appropriate for p-p collisions---is the assumption of FF universality inherent in the QCD factorization theorem relevant? The reply to 1) is currently no. The reply to 2) depends on context, as revealed by this analysis.

\subsection{The pQCD folding integral}

The folding integral used to obtain FDs in this analysis is
\bea \label{fold}
\frac{d^2n_{h}}{dy\, d\eta}  \hspace{-.05in} &\approx&  \frac{\epsilon(\delta \eta,\Delta \eta)}{ \sigma_{NSD}\, \Delta \eta}\int_0^\infty \hspace{-.07in}  dy_{max}\, D_\text{xx}(y,y_{max})\, \frac{d\sigma_{dijet}}{dy_{max}},
\eea
where $D_\text{xx}(y,y_{max})$ is the dijet FF ensemble  from a source collision system (xx = e-e, p-p, AA, in-medium or in-vacuum), and $d\sigma_{dijet}/dy_{max}$ is the minimum-bias parton spectrum. Spectrum hard component ${d^2n_{h}}/{dy\, d\eta}$ as defined represents the fragment yield from  scattered parton pairs into one unit of $\eta$. Efficiency factor $\epsilon \sim 0.5$ (for a single dijet and one unit of $\eta$) includes the probability that the second jet also falls within $\eta$ acceptance $\delta \eta$ and accounts for losses from jets near the acceptance boundary. $\Delta \eta \sim 5$ is the effective $4\pi$ $\eta$ interval for scattered partons. Further details are given in Sec.~\ref{jetspec}.


\subsection{Parton spectrum model} \label{parspec}


The {\em effective} parton spectrum for charged hadron fragments from p-p collisions can be estimated by folding a pQCD power-law parton spectrum hypothesis with trial FFs from p-p and e-e collisions and comparing the resulting FDs with the measured p-p spectrum hard component (interpreted as a fragment distribution).

A model for the parton $p_t$ spectrum resulting from minimum-bias scattering into an $\eta$ acceptance near projectile mid-rapidity can be parametrized as
\bea
\frac{1}{p_t}\,\frac{d\sigma_{dijet}}{dp_t} &=&   \frac{A_{p_t}}{p_t^{n_{QCD}}},
\eea
 which defines exponent $n_{QCD}$. The equivalent jet spectrum on $y_{max} \equiv\ln(2\,p_t / m_\pi )$ is
\bea
\frac{d\sigma_{dijet}}{dy_{max}} &=&  {A_{p_t}}\frac{p_t^2}{p_t^{n_{QCD}}} \\ \nonumber
&=& {A_{y_{max}}}\, \exp\{ -(n_{QCD} - 2)\,  y_{max}\},
\eea
where $p_t^2$ is the Jacobian factor for $p_t \rightarrow y_{max}$ and $y_{max} \geq y_{cut}$, the spectrum cutoff. The cutoff factor
\bea
 f_\text{cut}(y_{max}) = \{ \tanh[(y_{max} - y_{cut})/\xi_{cut}] + 1\}/2
\eea
represents in this analysis the minimum parton momentum which leads to detectable charged hadrons as neutral pairs (i.e., local charge ordering~\cite{ordering}).

\begin{figure}[h]
\includegraphics[height=.227\textwidth]{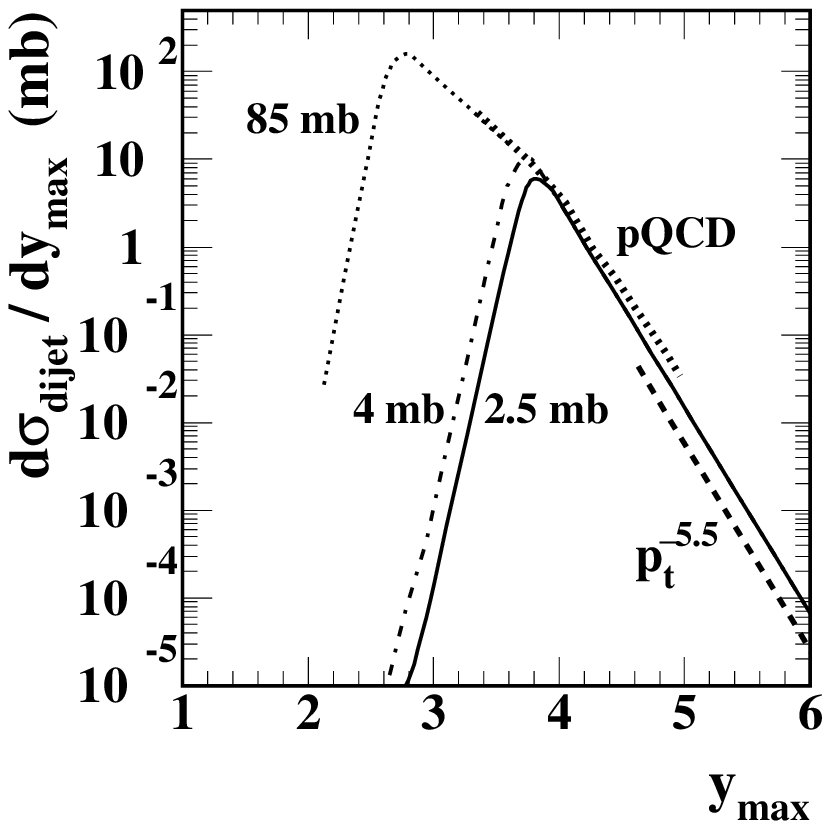}
\includegraphics[height=.23\textwidth]{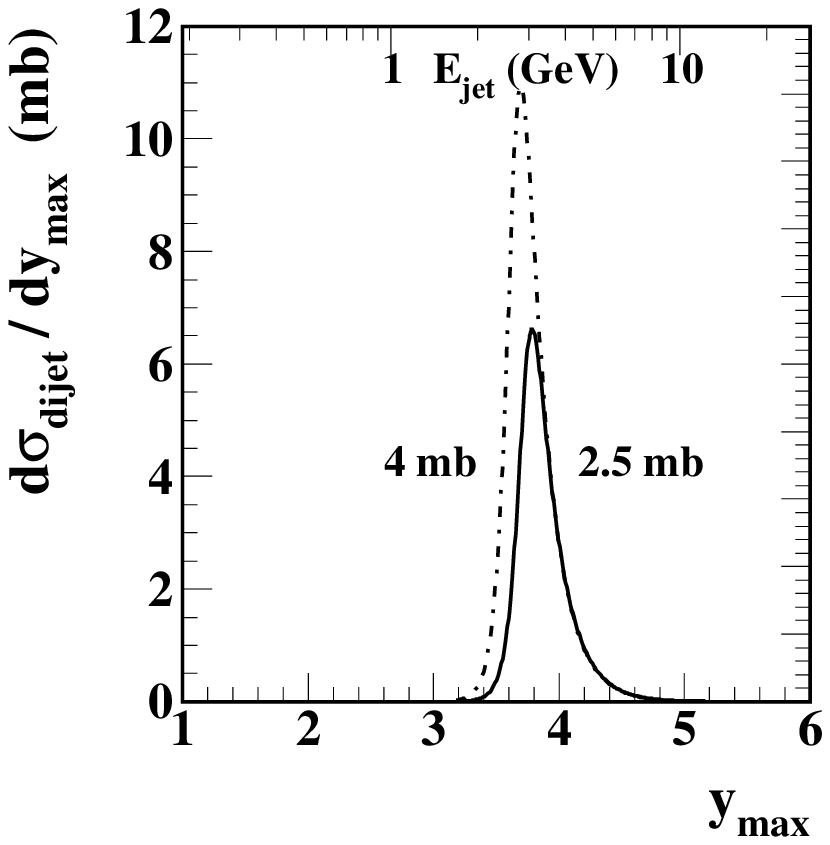}
\caption{\label{partspec} 
Dijet (parton-pair) transverse energy spectra on rapidity $y_{max} = \ln(2\, E_{jet} / m_\pi)$ plotted in semilog (left) and linear (right) formats. The solid curves are determined by a measured p-p spectrum hard component. The dash-dotted curves illustrate reduction of the cutoff energy inferred for central Au-Au collisions. The bold dotted curve labeled pQCD in the left panel is discussed in Sec.~\ref{disctheory}. The light dotted extrapolation down to 1 GeV corresponds to a saturation-scale cutoff estimate (Sec.~\ref{cgc}).
} 
\end{figure}

Fig.~\ref{partspec} (semilog and linear formats) shows a parton spectrum inferred from the p-p spectrum hard component in the next subsection. $y_{cut}$ is well-defined by the p-p hard component, and $n_{QCD}$ is defined by Au-Au spectrum hard components extending to larger $y_t$. Width parameter $\xi_{cut}$ affects details of FDs below the maximum (mode). Fixed value $\xi_{cut} = 0.1$ produces FD shapes consistent with $\xi_{cut} = 0$ but avoids a discontinuity. 

For a given value of jet cross section $\sigma_{dijet}$ the coefficient is $A_{y_{max}} = \sigma_{dijet}\, (n_{QCD} - 2)\, \exp\{(n_{QCD} - 2)\, y_{cut}\}$. For nominal values $n_{QCD} = 7.5$ and $y_{cut} = 3.75$ ($E_{cut} \sim 3$ GeV)  $\sigma_{dijet} \sim 2.5$ mb is adjusted to match FD data at larger $y$ (power-law tail), defining a fixed value of $A_{y_{max}}$. $y_{cut}$ is then adjusted to fit FD structure near the mode. The actual jet cross section varies {strongly} with $y_{cut}$ as
\bea
\frac{d\sigma_{dijet}}{\sigma_{dijet}} = -(n_{QCD} - 2)\, dy_{cut}.
\eea
A 0.1 reduction in $y_{cut}$, e.g.~from 3.75 to 3.65, (10\% relative reduction in $E_{cut}$) leads to a 55\% increase in the jet cross section (cf. Fig.~\ref{pspec} -- left panel). 

\subsection{Fragment distributions from p-p and e-e FFs}

\begin{figure}[h]
\includegraphics[width=.22\textwidth,height=.244\textwidth]{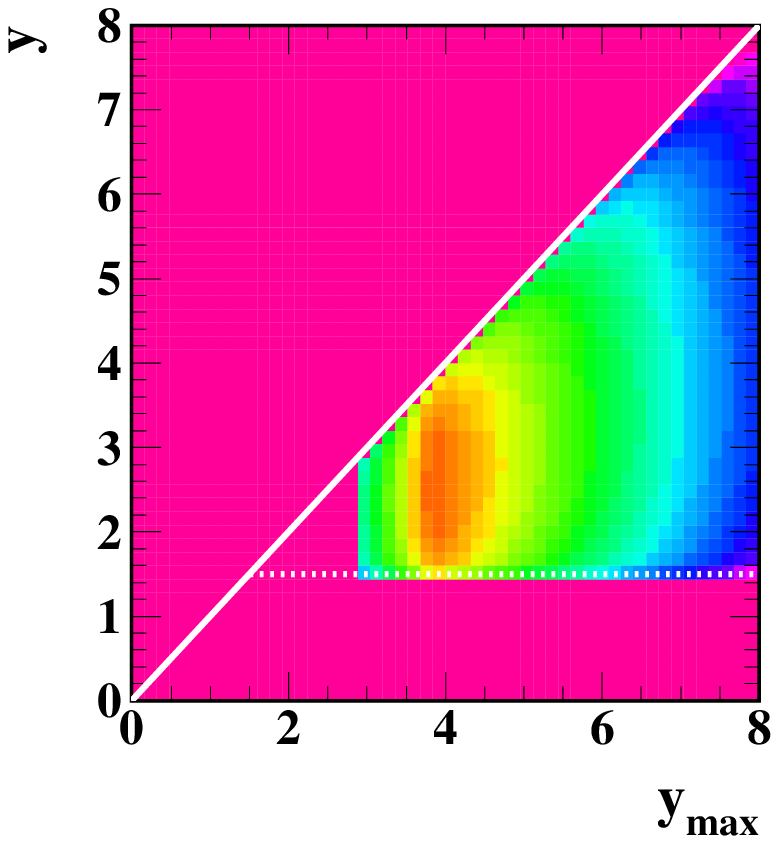}
\includegraphics[width=.22\textwidth,height=.24\textwidth]{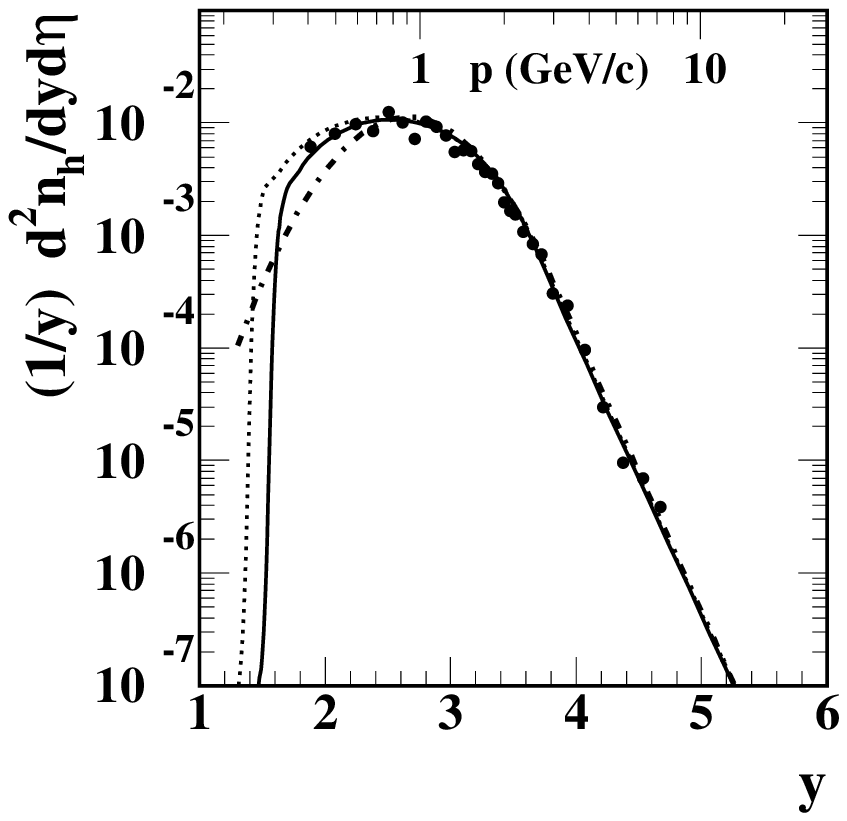}
\caption{\label{ppfd} 
Left: (Color online) Argument of the pQCD folding integral on $(y,y_{max})$ based on in-vacuum p-\=p FFs. 
Right: Fragment distribution $H_\text{NN-vac}$ (integral on $y_{max}$) obtained from in-vacuum p-\=p FFs (solid curve) compared to the Gaussian-plus-tail model of the p-p hard component (dash-dotted curve) and the measured hard component from NSD p-p collisions (solid points)~\cite{ppprd}. The dotted curve is discussed in Sec.~\ref{errors}.
} 
\end{figure}

Specific FF ensembles from LEP/HERA e-e and FNAL p-p collisions  can be combined with the parametrized parton spectrum to produce FDs for comparison with nuclear collision data. The hard component from p-p spectra determines the initial parton spectrum parameters for nuclear collisions. Fig.~\ref{ppfd} (left panel) shows a surface plot of the integrand of Eq.~(\ref{fold})---$D_\text{pp}(y,y_{max})\, \frac{d\sigma_{dijet}}{dy_{max}}$---incorporating FFs based on the LEP parametrization plus the FF cutoff inferred from p-\=p collisions. p-p FF distributions are bounded below by $y_{min} \sim 1.5$ ($p_t \sim 0.3$ GeV/c). The plot $z$ axis is logarithmic to show structure over the entire distribution support.

Fig.~\ref{ppfd} (right panel) shows the corresponding $H_\text{NN-vac}$ FD (integration of the left panel over $y_{max}$) as the solid curve. The mode of the FD is $\sim 1$ GeV/c. The dash-dotted curve is a Gaussian-plus-tail model function, and the solid points are hard-component data from p-p collisions~\cite{ppprd}. The comparison determines parton spectrum parameters $y_{cut} = 3.75$ ($E_{cut} \sim 3$ GeV) and exponent $n_{QCD} = 7.5$. The data are well-described by the pQCD folding integral.

The FD in the right panel represents the minimum-bias ensemble of jets that fall within the detector $\eta$ acceptance. 
The FDs are plotted as $(1/y)\, d^2n_{h} / dy\, d\eta$ for comparison to spectrum hard components plotted on transverse rapidity $y_t$.
The region above 2 GeV/c dominated by the pQCD power law is the conventional focus for study of parton fragmentation. Ironically, this analysis reveals that the physically most significant fragmentation structure and evolution lies below 2 GeV/c, the region conventionally assigned to hydro phenomena~\cite{2comp}.

\begin{figure}[h]
\includegraphics[width=.22\textwidth,height=.244\textwidth]{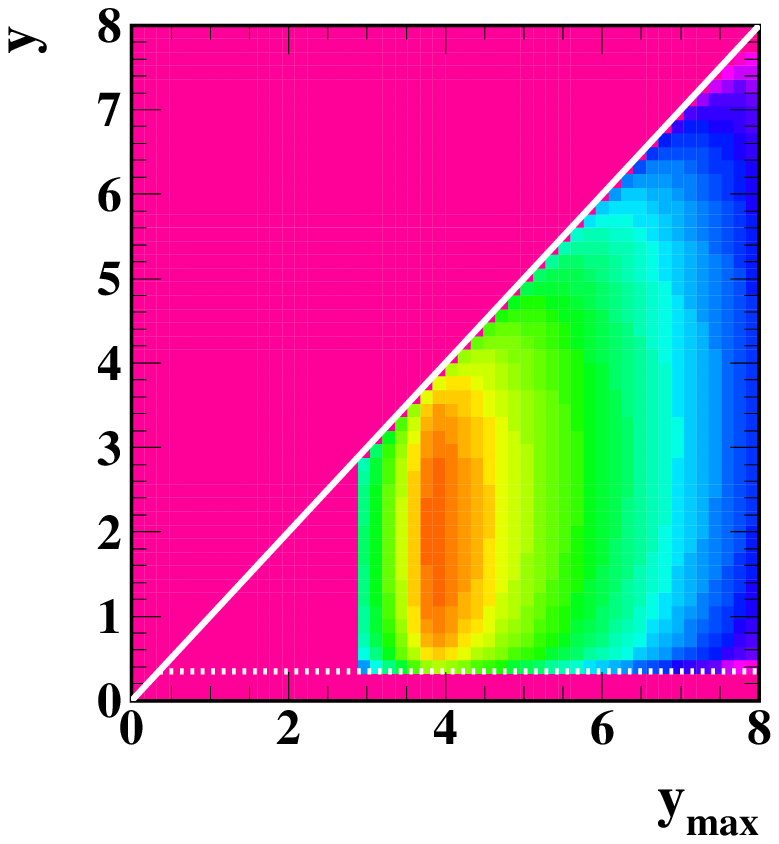}
\includegraphics[width=.22\textwidth,height=.24\textwidth]{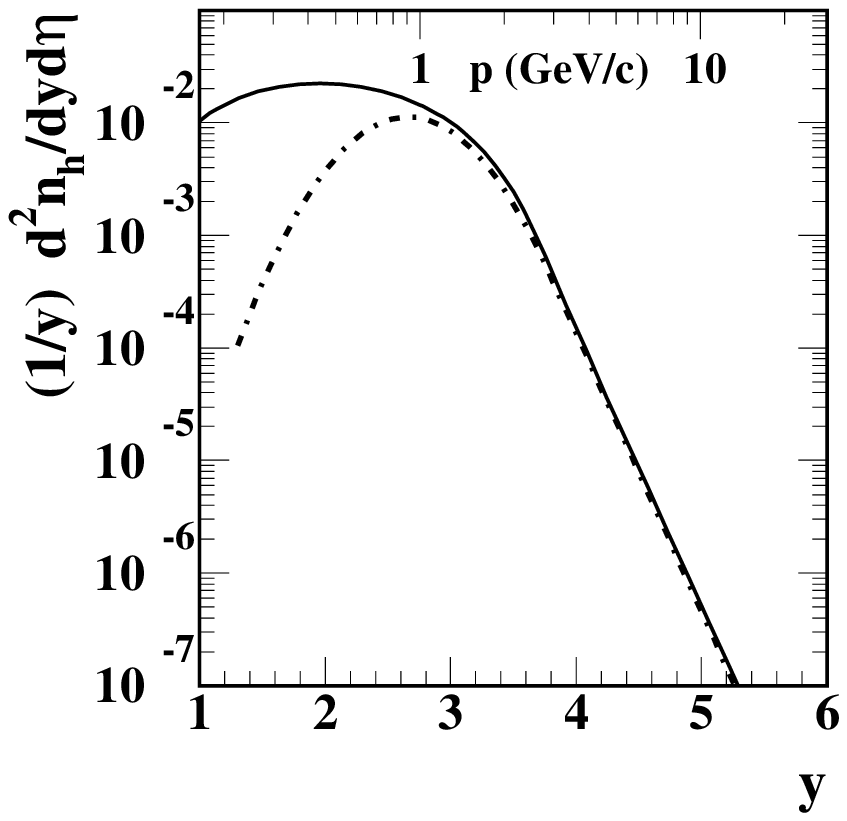}
\caption{\label{eefd} 
Left: (Color online) Argument of the pQCD folding integral on $(y,y_{max})$ based on in-vacuum $e^+$-$e^-$  FFs. 
Right: Fragment distribution $H_\text{ee-vac}$ (integral on $y_{max}$) obtained from in-vacuum $e^+$-$e^-$  FFs (solid curve) compared to the Gaussian-plus-tail model of the p-p hard component (dash-dotted curve)~\cite{ppprd}.
} 
\end{figure}

Fig.~\ref{eefd} (left panel) shows the argument of the folding integral incorporating unmodified FFs from e-e collisions. The main difference is the extension down to $y_{min} \sim 0.35$ ($p_t \sim 0.05$ GeV/c). 
Fig.~\ref{eefd} (right panel) shows  the corresponding FD (solid curve). The parton spectrum parameters determined by the p-p hard component are retained. The solid curve is the ``correct answer'' for an FD describing inclusive hadrons from inclusive partons produced by free parton scattering from p-p collisions, which is not observed in real nuclear collisions (cf. Sec.~\ref{revised}). The dash-dotted curve represents the hard-component model inferred from p-p collisions. The FD from e-e FFs lies well above the measured p-p hard component for hadron $p_t < 2$ GeV/c ($y_t < 3.3$), and the mode is reduced to $\sim 0.5$ GeV/c. The ``correct'' e-e FD strongly disagrees with the most relevant part of the p-p $p_t$ spectrum---the hard component. Despite strong disagreement the e-e FD is the correct reference for nuclear collisions, as shown below.

The p-p spectrum hard component determines the parton spectrum cutoff energy, the only adjustable parameter in the folding integral since the shapes of the p-p FFs are defined by independent FF data. The p-p FF multiplicity systematics in Fig.~\ref{ppffs2} suggest that the effective parton (gluon) spectrum cutoff is determined (for charge-particle measurements) by the requirement to produce at least one, and therefore two, charged hadrons. The cutoff inferred from p-p FFs ($y_{cut} \sim 3.75$) then provides an upper limit for \mbox{e-e} FFs, since the latter have substantially larger mean multiplicities. The caveat is especially relevant when modeling medium-modified FDs.

\subsection{NLO ``fragmentation functions''}

NLO FDs (sometimes termed ``fragmentation functions'')~\cite{nlo} are typically compared to the full p-p $p_t$ spectrum, including the soft as well as hard components~\cite{phenixnlo}. Theoretical representations of e-e FFs are currently less accurate for small fragment momenta (e.g., Fig.~20 of~\cite{ffprd}). As noted, inclusion of accurate in-vacuum e-e FFs in the pQCD folding integral leads to a large excess over the p-p hard component at smaller $y_t$, which may have important physical significance (as discussed below). The full soft+hard p-p spectrum greatly exceeds the FD at smaller $y_t$. Comparison of NLO FDs with the full p-p spectrum can then suggest agreement with data which is misleading. Instead, they should be compared directly to the p-p spectrum hard component as in Fig.~\ref{eefd} (right panel), which then reveals physically important differences.

\section{The parton spectrum} \label{jetspec}

The $p_t$ spectrum for partons scattered from N-N collisions can be approximated by $(1/p_t)\, d\sigma_{dijet}/ dp_t = A\, p_t^{-n_{QCD}}$ above some cutoff $p_{t,cut}$. The three  spectrum constants ($A$, $n_{QCD}$, $p_{t,cut}$) can be inferred from nuclear collision data. Parton spectrum information comes from event-wise jet reconstruction and from the single-particle spectrum hard component. Jet reconstruction provides only a part of the differential spectrum. $A$ and $n_{QCD}$ can be inferred from the larger-$p_t$ region. A $p_t$-spectrum hard component combined with an FF ensemble can determine the parton spectrum cutoff explicitly from the structure near its mode. The parton (dijet) total cross section is then determined with improved accuracy.

\subsection{Parton spectrum from data} \label{specdat}

When integrated over fragment rapidity $y$ the folding integral in Eq.~(\ref{fold}) can be expressed as the product of a weighted-mean dijet multiplicity and the integrated dijet cross section. Since the parton spectrum is sharply peaked near 3 GeV the mean dijet multiplicity (Fig.~\ref{ppffs2} -- right panel) is close to the value at that energy. Factor $\epsilon$ is introduced to represent the 1D dijet efficiency: the average fraction of a jet that falls inside $\eta$ acceptance bin $\delta \eta$ (some jets overlap the boundary) and the fraction of partners in a back-to-back jet pair that also fall inside $\delta \eta$. Both fractions depend on $\delta \eta$, effective 4$\pi$ interval $\Delta \eta$ and the number $N_{dijet}$ of jet pairs in a nuclear collision.

\begin{figure}[h]
\includegraphics[height=.225\textwidth]{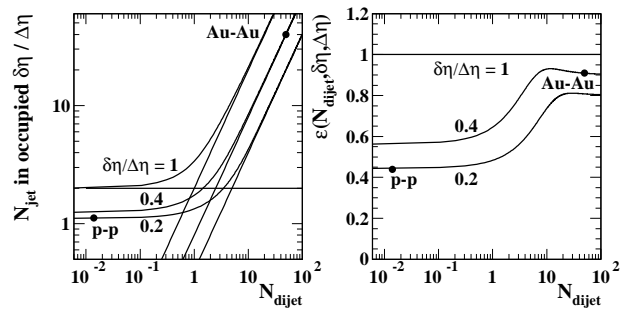}
\caption{\label{jetfrac}  
Left: Mean jet number $N_{jet}$ in an {\em occupied} bin $\delta \eta$ within acceptance $\Delta \eta$ for dijet number $N_{dijet}$ and assuming Poisson statistics.
Right: Fraction $\epsilon$ of dijet fragment yield in occupied bin $\delta \eta$ within interval $\Delta \eta$ for dijet number $N_{dijet}$.
} 
\end{figure}

The $y_t$-integrated two-component model of mean hadron yields for NSD p-p collisions is
\bea
\,\frac{dn_{ch}}{d\eta} &=& \frac{dn_s}{d\eta} + \frac{dn_h}{d\eta}.
\eea
The jet cross section can be inferred from measurements of spectrum hard component $H = (1/y)\, d^2 n_h / dy\, d\eta$. Integrating Eq.~(\ref{fold}) over y and expressing the remaining $y_{max}$ integral as $\bar n_{dijet} \, {\sigma_{dijet}}$ gives
\bea \label{jetcross}
\frac{dn_h}{d\eta} &\approx& \left\{ \frac{1}{\sigma_{NSD}}
\, \frac{\sigma_{dijet}}{\Delta \eta}\right\} {\epsilon(\delta \eta,\Delta \eta)\, \bar n_{dijet} }
\eea
for NSD p-p collisions. The measured hard/soft ratio for NSD mean $\hat n_{ch} \sim 1.25$ is $n_h/ n_s \sim 0.008$~\cite{ppprd}. Since $dn_s/d\eta \sim 2.5$ for NSD p-p collisions $dn_h/d\eta \sim 0.02$. Given $\sigma_{NSD} \sim 36$ mb, p-\=p $\bar n_{dijet} \sim 3$ and  dijet fraction  $\epsilon \sim 0.45$ (cf. Fig.~\ref{jetfrac}) included in $|\eta| < 0.5$ ($\delta \eta = 1$) then $\sigma_{dijet} / \Delta \eta = 0.5 \pm 0.12$ mb. Assuming $\delta \eta_{4\pi} \equiv \Delta \eta \sim 5$ gives a total cross section $\sigma_{dijet} = 2.5 \pm 0.6$ mb. Uncertainty estimates are discussed in Sec.~\ref{errors}.
Equivalently, the probability of a minijet within $|\eta| < 0.5$ in NSD p-p collisions [expression within curly brackets in Eq.~(\ref{jetcross})] is 0.5 mb/36 mb $\sim 0.014\pm 0.003$, consistent with~\cite{ppprd} based on $\epsilon\, \bar n_{dijet} \sim 2.5\pm 1.0$ and $dn_h/d\eta \sim \alpha\, \hat n_{ch}\, dn_{ch}/d\eta \sim 0.03\pm 0.01$.

The correspondence between a dijet cross section and average jet fragment yield in an $\eta$ acceptance is nontrivial.  The general result for A-A collisions depends on mean dijet number $N_{dijet} = n_{binary}\, \sigma_{dijet} / \sigma_{NSD}$. pQCD calculations produce differential cross section $d^3\sigma_{dijet}/dp_t\, dy_1\, dy_2$, with $d\sigma_{dijet}/dp_t$ or $d\sigma_{dijet}/dy_{max}$ as a straightforward 1D projection. To obtain a 2D projection onto $y_{max}$ and $y$ or $\eta$ the single integral over $y$ must accommodate the integer number of dijets in a collision. 

Fig.~\ref{jetfrac} (left panel) shows calculated mean jet multiplicities $N_{jet}$ for {\em occupied} bins of width $\delta \eta$ within $4\pi$ acceptance $\Delta \eta$. $N_{jet}$ varies as $2\, N_{dijet}\, \delta \eta / \Delta \eta$ toward the right. Toward the left where $N_{dijet} \ll 1$ the mean jet number in an occupied bin is $1/(1-0.5\,\delta \eta / \Delta \eta)$.

In Fig.~\ref{jetfrac} (right panel) $\epsilon(N_{dijet},\delta \eta,\Delta \eta)$ represents the fractional yield of hadrons per dijet into acceptance $\delta \eta$. For a single dijet in $\Delta \eta \sim 5$ (e.g., some p-p collisions) and $\delta \eta = 1$,  $\epsilon(1,1,5) \sim 0.8/(2 - \delta \eta / \Delta \eta) \sim 0.45$. For $N_{dijet} \sim 50$ (e.g., semi-central Au-Au collisions) and $\delta \eta = 2$, $\epsilon(50,2,5)  \sim 0.9$, a two-fold increase in fractional hadron yield per dijet.


\begin{figure}[h]
\includegraphics[height=.225\textwidth]{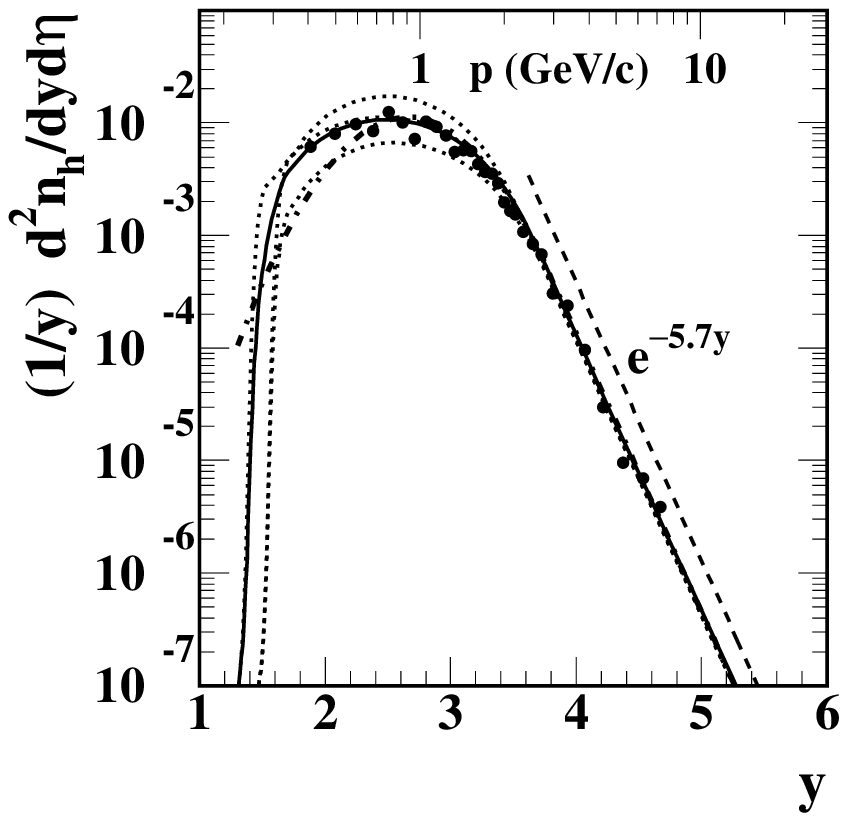}
\includegraphics[height=.225\textwidth]{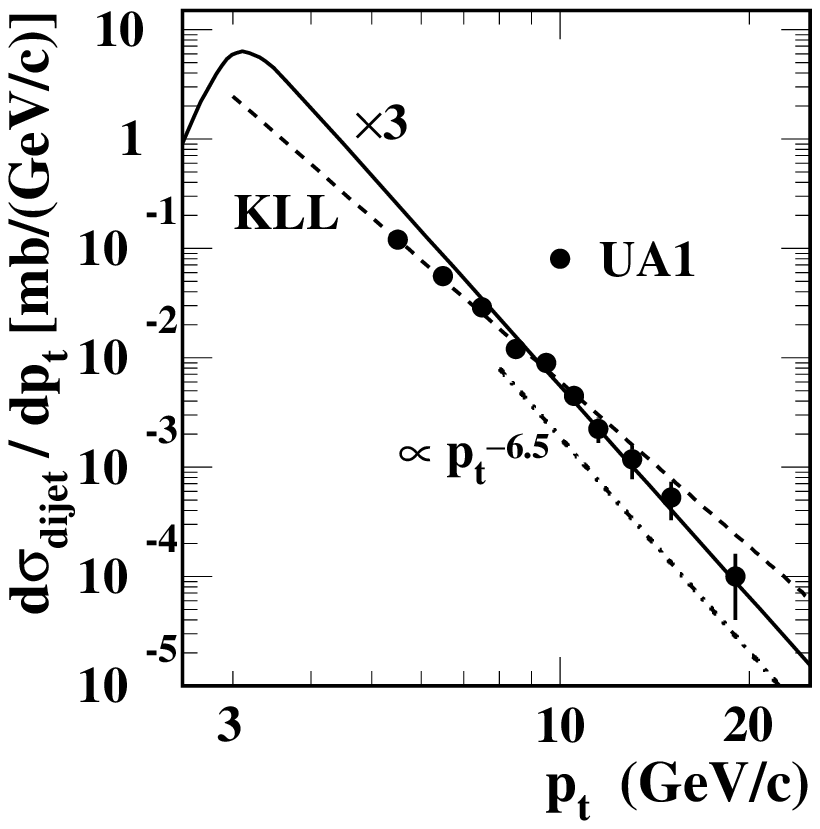}
\caption{\label{pspec}  
Left: $H_\text{NN-vac}$ for a nominal 3 GeV spectrum cutoff (solid curve) and $y_{cut}$ shifted by 0.1 ($E_{cut}$ changed by 10\%) in either direction (dotted curves) compared to p-p reference $H_\text{GG}$ (dash-dotted curve) and NSD p-p data (solid points), illustrating the precision of the parton spectrum cutoff determination.
Right: The parton spectrum defined in this analysis with $n_{QCD} \sim 7.5$ integrating to $\sim 2.5$ mb multiplied by $3$ (solid curve), KLL parametrized spectrum~\cite{kll} integrating to 2.2 mb above 3 GeV/c (dashed curve) and UA1 measured jet differential cross section integrating to $\sim 4$ mb (points)~\cite{ua1}.
} 
\end{figure}

Figure~\ref{pspec} (left panel) shows the calculated p-p FD for spectrum cutoff $y_{cut} = 3.75$ (solid curve) and for $y_{cut} =$ 3.65, 3.85 corresponding to 10\% shifts in the cut energy about 3 GeV (dotted curves) illustrating the precision of the cutoff determined from data: p-p data determine the cutoff to better than 5\%. 

\subsection{Parton spectrum from theory}

Figure~\ref{pspec} (right panel) compares the spectrum defined by this analysis (solid curve, and note the factor 3) with theory and 200 GeV UA1 data. As noted, the spectrum from this analysis integrates to $2.5 \pm 0.6$ mb with well-defined cutoff $\sim 3$ GeV. The KLL parametrization $600/p_t^5$ mb/(GeV/c)~\cite{kll} (dashed line) integrates to 2.2 mb above 3 GeV/c. 

The UA1 spectrum (points) is comparable at larger $p_t$ with the spectrum from this analysis (multiplied by 3) in an interval where the jet-finding efficiency should be good and underlying-event contributions are relatively small. The slopes are the same but the amplitudes are different. At smaller $p_t$ the UA1 data fall below the solid curve in a region where jet-finding efficiency might be reduced. The UA1 spectrum integral is 4 mb~\cite{ua1}.

Several theoretical calculations of the jet total cross section stimulated by the UA1 minijet data were intended to extend pQCD comparisons with data down to $\sim3$ GeV. In \cite{panch} a ``two-component'' spectrum model for p-p collisions was discussed, and it was estimated that because of the underlying event the $E_t \sim 5$ GeV observed by UA1 corresponds to 3-4 GeV/c parton momentum. The total cross section obtained from a pQCD calculation was 2-2.5 mb for a 3 GeV spectrum cutoff.

A similar calculation in~\cite{durand} obtained 3 mb for a jet energy threshold of 3 GeV. However, the assertion that there is no actual threshold for parton $\rightarrow$ hadron is contradicted for {\em charged} hadrons by the present study. The mean number of jets in NSD p-p collisions was given as 
$2 \sigma_{dijet} / \sigma_{inelastic \rightarrow NSD} \sim 0.1$-0.2, which compares with $2 \times 2.5\pm0.6$ mb/ 36 mb = $0.14\pm0.03$ from this analysis.

An extensive study of effects on the minijet cross section from energy scaling of structure functions was described in~\cite{sarc}. The goal was to apply pQCD to low-$p_t$ jet physics---``the so-called minijet regime.'' Estimation of the underlying-event contribution to the UA1 $E_t$ cluster finder lead to the conclusion that a 5 GeV $E_t$ calorimeter cluster (minijet) translates to a 3 GeV parton. Almost all partons in the spectrum (i.e., with energies $\sim 3$ GeV) are gluons. The minijet cross section was estimated to be 2-3 mb at 200 GeV, with spectrum cutoff near 3 GeV.



Based on comparisons of spectrum hard components with calculated FDs and with pQCD theory there can be considerable confidence that the measured spectrum hard component from p-p collisions is a parton fragment distribution, and the corresponding minimum-bias parton spectrum is well defined. Given that baseline we now consider parton ``energy loss'' in A-A collisions.

\section{Parton ``energy loss''}

In conventional pQCD descriptions of parton energy loss the leading parton is said to lose energy by gluon bremsstrahlung during passage through a (possibly colored) medium~\cite{parteloss}. In the context of this analysis some questions emerge: 1) Do {\em complete} FDs in more-central A-A collisions actually indicate energy loss (i.e., do FFs integrate to reduced parton energy)? 2) Do jet angular correlations manifest structure changes consistent with leading-parton random multiple scattering (e.g., broadening {\em symmetric} about the jet axis)? 3) If the answer to 1) is ``yes'' how and where is the lost energy manifested in the medium? 4) If the answer to 1) or 2) is ``no'' what is the relevance or proof of an independent medium? To address those questions we can incorporate energy-loss models into calculated FDs from e-e and p-p collisions and compare with hard components (single-particle spectra {and correlations}) measured in Au-Au collisions.

\subsection{Negative boost of the p-p hard component}

A simple algebraic model of energy loss in FDs can be obtained by shifting p-p or N-N hard-component model $H_{GG}$ down on rapidity (negative boost $\Delta y_t \sim \Delta E _{parton}/ E_{parton}$). The model manifestly does not conserve energy (energy $\Delta E$ is lost from the fragment system). Ratio $r_{AA}$~\cite{2comp} is then modeled as $\ln(r_{AA})$ by
\bea
\ln\left\{ \frac{H_{GG}(y_t+\Delta y_t)}{H_{GG}(y_t)} \right\}  \approx  -\Delta y_t \cdot \frac{d\ln(H_{GG})}{dy_t},
\eea
where $H_{GG}$ is the Gaussian plus power-law tail inferred from p-p collisions (denoted by $n_h\, H_0$ in~\cite{ppprd}). The $H_{GG}$ reference is included in plots of $r_{AA}$ below. The negative-boost model is imperfect because it does not respect the details of QCD splitting, and the lower limit of the FD experiences the same rapidity shift as the leading parton, which is probably not physical (cf. the next subsection). Its recommendation is algebraic simplicity.

\subsection{``Medium-modified'' FFs} \label{bweloss}

A better-justified model (BW) of medium-induced modification to QCD fragmentation is described in~\cite{b-w}. Parton ``energy loss'' is modeled, but the parton energy is conserved within the modified FF. In contrast to special treatment of the leading parton (bremsstrahlung) with ``broadening and softening'' of the FF, all subleading splittings are treated equally by BW---momentum is conserved at all stages of the cascade. 

The BW model is applied to MLLA descriptions of two experimental FFs (TASSO 14 GeV~\cite{tasso} and OPAL 200 GeV~\cite{opal200}). As noted in~\cite{ffprd} MLLA FF parametrizations fail for small and large fragment momenta. The bottom 20\% of the FF, where the most interesting conclusions of this analysis emerge, is typically missing. Discrepancies are typically of the same magnitude as the medium effects observed in this analysis using an accurate FF representation. The statement ``the MLLA can serve as a baseline in searching for medium effects'' is not justified. 

\begin{figure}[h]
   \includegraphics[width=1.65in,height=1.65in]{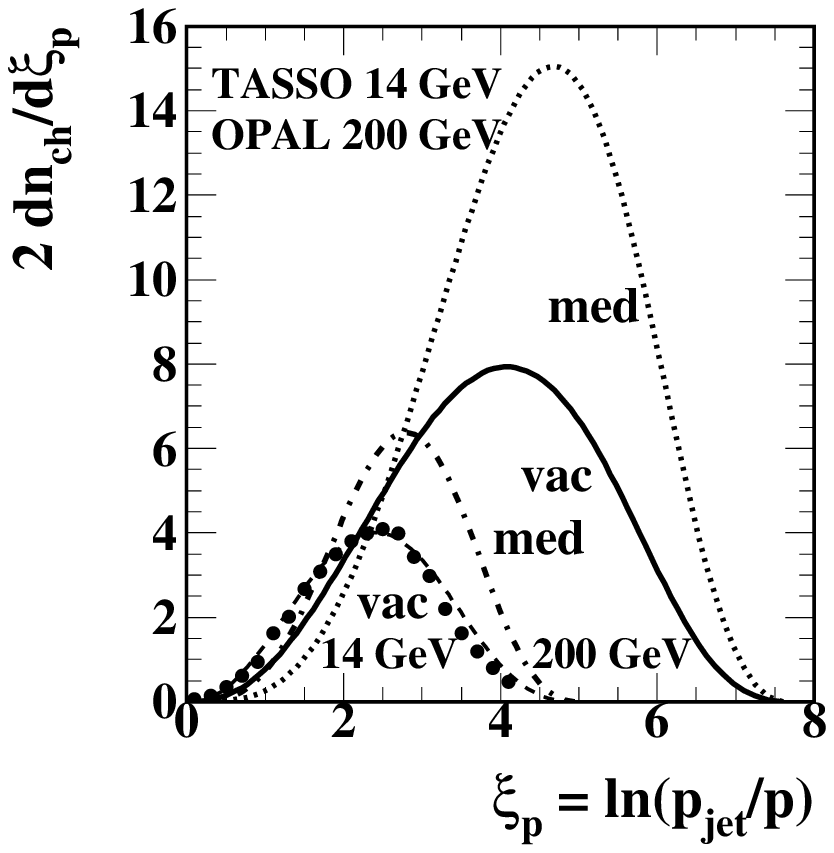}
 \includegraphics[width=1.65in,height=1.65in]{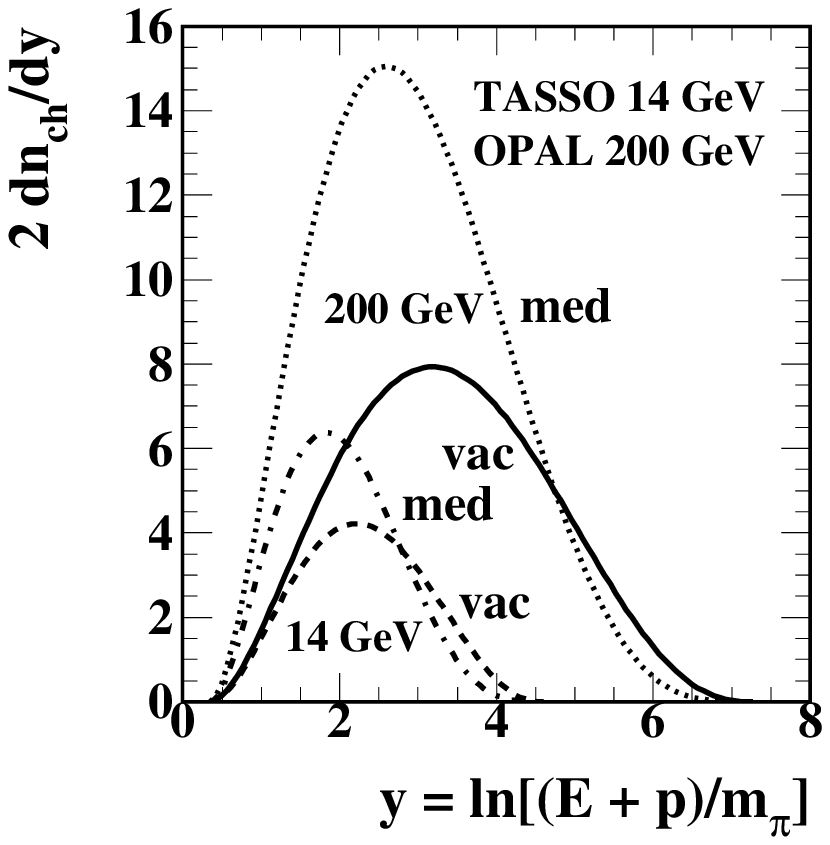}
\caption{\label{eloss1}
 Left: parametrized fragmentation functions from $e^+$-$e^-$ collisions at two energies plotted on $\xi_p$ for in-vacuum  FFs (dashed and solid curves)~\cite{ffprd} and for in-medium modification by rescaled splitting functions (dash-dotted and dotted curves)~\cite{b-w} compared to data for 14 GeV (solid points)~\cite{tasso}.
Right: Curves in the left panel replotted on rapidity $y$.
} 
 \end{figure}

Figure~\ref{eloss1} (left panel) illustrates the BW model (cf. Fig. 1 of~\cite{b-w}). In-vacuum FFs for $Q = 14$ and 200 GeV derived from the beta parametrization are shown as the dashed and solid curves respectively~\cite{ffprd}. Data from TASSO 14 GeV (solid points) are shown for comparison, duplicating part of Fig.~\ref{eeffs} (left panel). The beta distribution represents all FF data to the statistical limits. The practical consequence of the BW ``energy-loss'' mechanism is a momentum-conserving rescaling of FFs on momentum fraction $x_p$ or logarithmic variable $\xi_p = \ln(1/x_p)$. Density reductions at larger fragment momenta (smaller $\xi_p$) are balanced by much larger increases at smaller momenta. The large changes correspond to an inferred leading-parton fractional ``energy loss'' of  25\%.

The BW model modifies the splitting process by rescaling the momentum fraction consistently at all stages of the cascade. The beta parametrization of FFs~\cite{ffprd} has only  two parameters $(p,q)$. $p$ represents the effect of quantum coherence or gluon saturation which effectively terminates the cascade at hadron formation (LPHD). $q$ represents the pQCD splitting cascade itself. Empirically, we observe that increasing parameter $q$ by $O(1)$ increment $\Delta q$ accurately duplicates the BW rescaling process. 

The dotted and dash-dotted curves in Fig.~\ref{eloss1} correspond to $q \rightarrow q + 1.15$. The curves (left panel) match Fig. 1 of~\cite{b-w} in the interval for which the MLLA approximates FF data. We can thus implement the BW energy-loss model simply by changing $q$ in the beta parametrization to achieve an accurate and complete representation of measured FFs. Figure~\ref{eeffs2} (right panel) uses the beta parametrization to demonstrate the multiplicity increase (open circles) corresponding to the BW ``energy-loss'' (medium modification) prescription.

Figure~\ref{eloss1} (right panel) shows the same system on fragment rapidity $y$. The relation is given by  $\xi_p = \ln(p_{jet}/ p)$ = $\ln(2\, p_{jet} / m_\pi) - \ln(2p/m_\pi) \sim y_{max} - y$, with energy scale $Q = 2\, p_{jet}$. Beta FFs on $\xi_p$ don't extend to infinity because FFs on $y$ are bounded below by $y_{min}$. In e-e collisions $y_{min} \sim 0.35$ corresponds to  $p \sim m_\pi / 2$. The maximum $\xi_p$ value is thus $\xi_{p,max} \sim \ln(Q/m_\pi) = y_{max}$ ($\sim \tau$ in~\cite{b-w}).

The BW ``energy-loss'' model coupled with the FF beta parametrization from~\cite{ffprd} provides an accurate algebraic model of FF ``medium modification'' valid for all relevant energies and momenta and directly related to pQCD principles. The system can generate medium-modified FDs for comparison with A-A hard-component evolution.


\subsection{``Energy loss'' and FDs}

\begin{figure}[h]
\includegraphics[width=1.65in,height=1.77in]{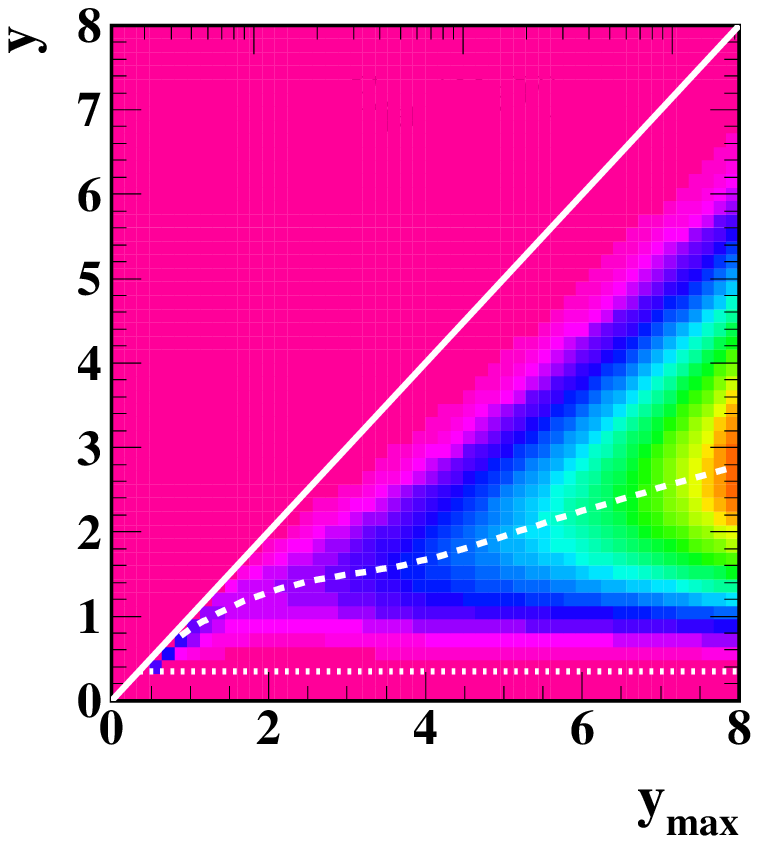}
\includegraphics[width=1.65in,height=1.75in]{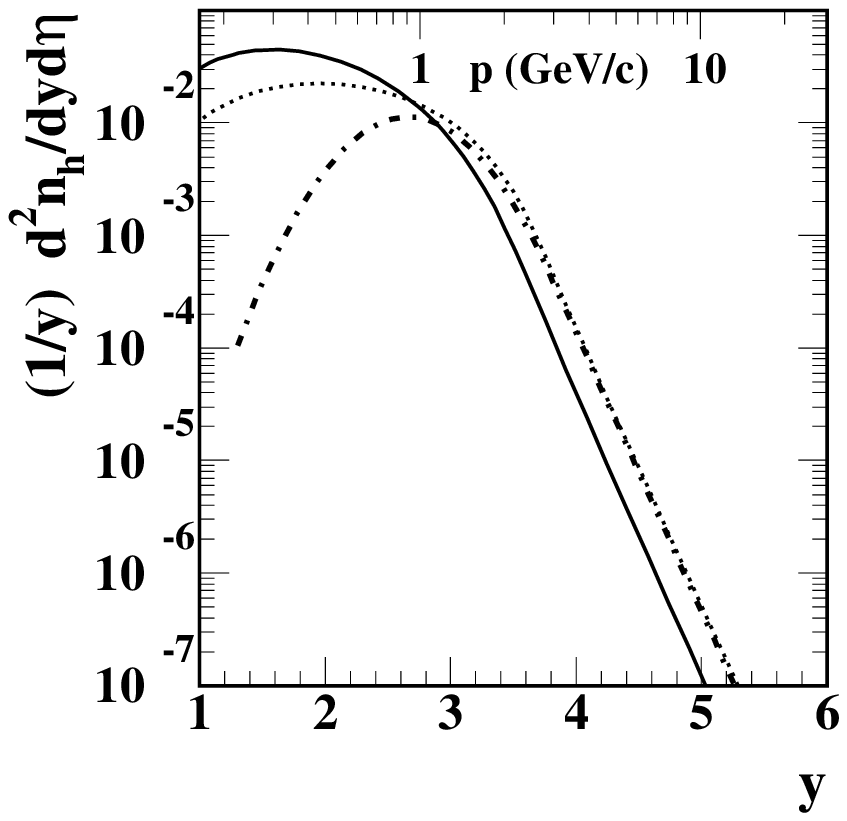}
\caption{\label{eloss2} 
Left: (Color online) Argument of the pQCD folding integral on $(y,y_{max})$ based on in-medium $e^+$-$e^-$  FFs. 
Right: Fragment distribution $H_\text{ee-med}$ (integral on $y_{max}$) obtained from in-medium $e^+$-$e^-$  FFs (solid curve) compared to the in-vacuum FD (dotted curve) and the Gaussian-plus-tail model of the p-p hard component (dash-dotted curve)~\cite{ppprd}.
} 
\end{figure}

Figure~\ref{eloss2} (left panel) shows the e-e FF ensemble with BW modification as described in the previous subsection. Energy-loss parameter $\Delta q$ is the change in beta-distribution model parameter $q$ which emulates the BW energy-loss method. The value $\Delta q = 1.15$ (for 0-12\% central Au-Au collisions) is determined by hard-component data above $p_t \sim 4$ GeV/c, where ``suppression'' in ratio $r_{AA}$ is approximately constant. For this initial survey $\Delta q$ is assumed to be independent of parton energy ($y_{max}$).

Figure~\ref{eloss2} (right panel) shows $H_\text{ee-med}$ (solid curve), the FD obtained by inserting e-e in-medium FFs from the left panel into Eq.~(\ref{fold}) and integrating over parton rapidity $y_{max}$. The dotted curve is the $H_\text{ee-vac}$ reference from in-vacuum e-e FFs. The dash-dotted curve is again the Gaussian-plus-tail p-p hard component $H_{GG}$ for reference. The mode of $H_\text{ee-med}$ is $\sim 0.3$ GeV/c.

\begin{figure}[h]
\includegraphics[width=1.65in,height=1.77in]{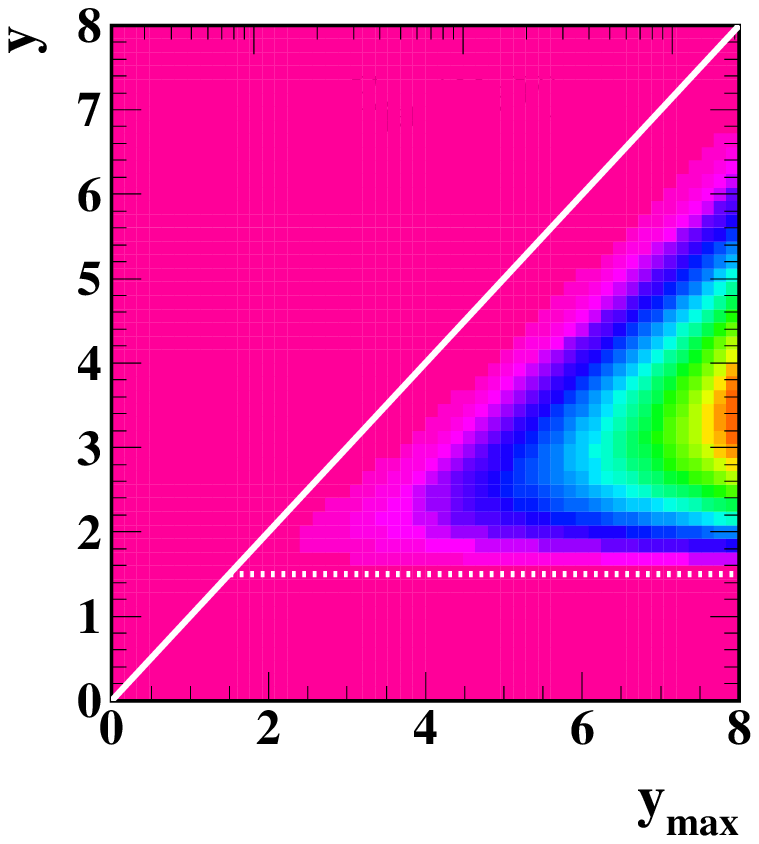}
\includegraphics[width=1.65in,height=1.75in]{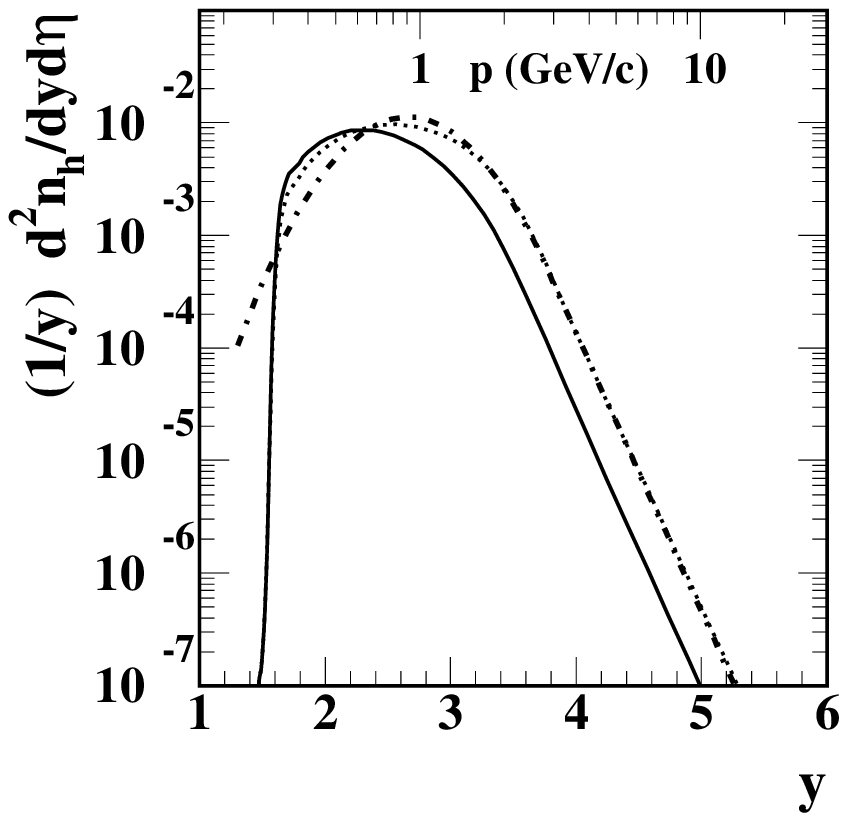}
\caption{\label{eloss3} 
Left: (Color online) Argument of the pQCD folding integral on $(y,y_{max})$ based on in-medium p-\=p  FFs. 
Right: Fragment distribution $H_\text{NN-med}$ (integral on $y_{max}$) obtained from in-medium p-\=p  FFs (solid curve) compared to the in-vacuum FD (dotted curve) and the Gaussian-plus-tail model of the p-p hard component (dash-dotted curve)~\cite{ppprd}.
} 
\end{figure}

Figure~\ref{eloss3} shows results for p-p FFs. The major difference between p-p and e-e FDs appears below $p_t \sim 2$ GeV/c ($y_t \sim 3.3$). Conventional comparisons with theory (e.g., data {\em vs} NLO FDs) typically do not extend below 2 GeV/c~\cite{phenixnlo}. 

\section{A-A two-component spectra}

This analysis describes spectrum hard components by folding a parton power-law spectrum with parametrized FF ensembles. The hard component from p-p collisions constrains the parton spectrum. Hard components have also been extracted from Au-Au spectra for five centralities~\cite{2comp}. A sharp transition in hard-component properties is observed as for minijet angular correlations~\cite{daugherity}. We use measured Au-Au hard components to study the evolution of parton fragmentation (and possibly the underlying parton spectrum) with centrality in nuclear collisions.

\subsection{A-A two-component model}

The algebraic form of the two-component model of per-participant-pair A-A spectra is 
\bea  \label{aa2comp}
\frac{2}{n_{part}} \frac{1}{y_t}\frac{dn_{ch}}{dy_t} &=& S_{NN}(y_t) +  \nu\, H_{AA}(y_t;\nu) \\ \nonumber
&=&  S_{NN}(y_t) +  \nu\,r_{AA}(y_t;\nu) \,H_{NN}(y_t),
\eea
where $S_{NN}$ ($\sim S_{pp}$) is the soft component and $H_{AA}$ is the hard component (with reference $H_{NN}\sim H_{pp}$) integrating respectively to multiplicities $n_s$ and $n_h$ in one unit of pseudorapidity $\eta$~\cite{ppprd,2comp}. Ratio $r_{AA} = H_{AA} / H_{NN}$ is a refinement of nuclear modification factor $R_{AA}$. Centrality measure $\nu \equiv 2 n_{binary} / n_{participant}$ estimates a mean participant-nucleon path length in the Glauber model. Model functions $S_{xx}$ and $H_{xx}$ are normalized to be compatible with the total spectrum density in a given context (i.e., whether the density is 1D, 2D or 3D).

The soft component is interpreted as longitudinal projectile fragmentation ({\em via} inelastic N-N scattering) approximately independent of A-A centrality. The hard component is interpreted (as in p-p collisions) as the FD from minimum-bias parton scattering and fragmentation into some hadron angular acceptance. Because a single soft-component model function is subtracted from all centralities any systematic error in the subtraction is common to all hard components. {\em Relative} variations with centrality are then unique to hard-component structure.

\subsection{Hard-component spectra and ratios}

\begin{figure}[h]
 \includegraphics[width=1.65in,height=1.67in]{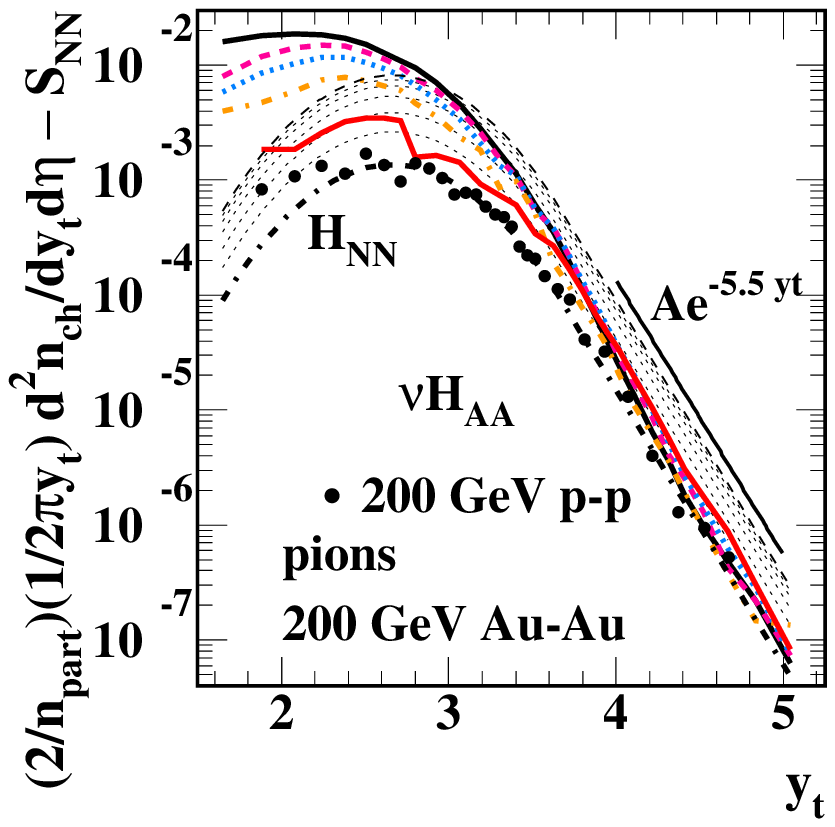}
 \includegraphics[width=1.65in]{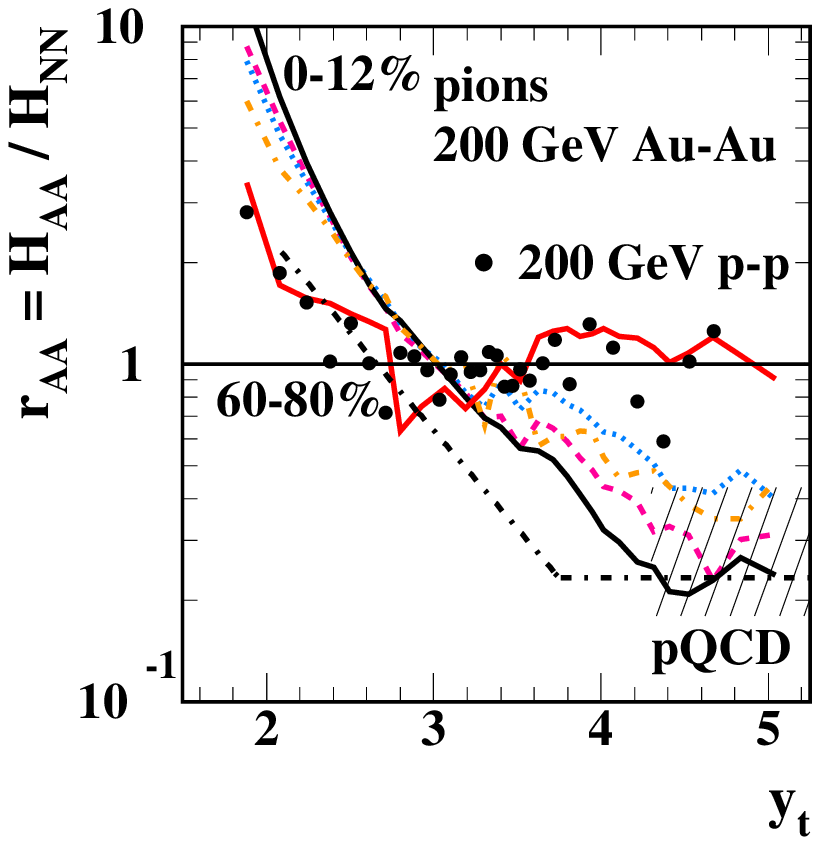}
\caption{\label{aaspec} (Color online)
 Left: Spectrum hard components from five centralities of 200 GeV Au-Au collisions (bold curves of several types)~\cite{2comp} compared to a two-component reference system (thin dotted curves). Hard-component data from p-p collisions (solid points)~\cite{ppprd} are included for comparison.
Right: Hard-component ratio $r_{AA}$ for the data in the left panel (bold curves of several types and dots) compared to a simple energy-loss model (black dash-dotted curve).
} 
\end{figure}

Figure~\ref{aaspec} (left panel) shows hard-component evolution with centrality for pions from 200 GeV Au-Au collisions (five centrality classes). The spectrum data are in the form of a 3D density on $(y_t,\eta,\phi)$. The thin dotted reference curves are obtained from Eq.~(\ref{aa2comp}) by replacing $H_{AA}$ with reference $H_{NN}$ (model function $H_{GG}$). The points are the hard component from 200 GeV NSD p-p collisions~\cite{ppprd}. The main features are the suppression at larger $y_t$ intensively studied at RHIC and the much larger enhancement at smaller $y_t$ first described in~\cite{2comp}.

Figure~\ref{aaspec} (right panel) shows corresponding ratio $r_{AA}$ based on hard-component reference $H_{NN}$ set equal to Gaussian model $H_{GG} = n_h\, H_0$ from~\cite{ppprd}. Evolution of suppressions and enhancements is more clearly visible. The p-p data and the most peripheral Au-Au data agree with the N-N reference ($r_{AA} = 1$) above $y_t = 2.5$ but deviate significantly from $H_{GG}$ below that point.

As noted in~\cite{2comp} there is a sharp transition in the centrality trend for both suppression and enhancement, also seen in 200 GeV Au-Au angular correlation studies~\cite{daugherity}. The direct correspondence between the centrality trend at $p_t \sim 10$ GeV/c and that at $p_t \sim$ 0.3 GeV/c seems curious in the conventional RHIC context (hydro + high-$p_t$ jets). However, if the entire hard component is interpreted as an FD the correlation is seen to be inevitable. 

A major consequence of this analysis is the realization that $H_{NN}$ from p-p collisions is not the correct FD reference for nuclear collisions. The p-p hard component is itself strongly modified relative to the correct reference. An alternative to $r_{AA}$ is required for differential study.

\section{Centrality evolution of the FD} \label{centrality}

The insensitivity of nuclear modification factor $R_{AA}$ to most energy-loss details and the superiority of hard-component (FD) ratio $r_{AA} = H_{AA} / H_{NN}$ were demonstrated in~\cite{2comp}. In this section calculated FDs are compared directly with measured A-A spectrum hard components using generalized hard-component ratio $r_{xx}$ to determine centrality evolution. The notation adopted is FD $\rightarrow H_\text{xx}$, where xx = pp (data), AA (data), GG (Gaussian-plus-tail model) and NN-vac, NN-med, ee-vac, ee-med. The last four, introduced in the previous section, are obtained from folding integrals. We then define ratios $r_\text{AA} = H_\text{AA} / H_\text{GG}$ (used in~\cite{2comp} and Fig.~\ref{aaspec} -- right panel),  $r_\text{ee} = H_\text{ee-med} / H_\text{ee-vac}$ and  $r_\text{NN} = H_\text{NN-med} / H_\text{NN-vac}$.

\subsection{FD ratios compared to central Au-Au collisions}

\begin{figure}[h]
\includegraphics[width=1.65in,height=1.75in]{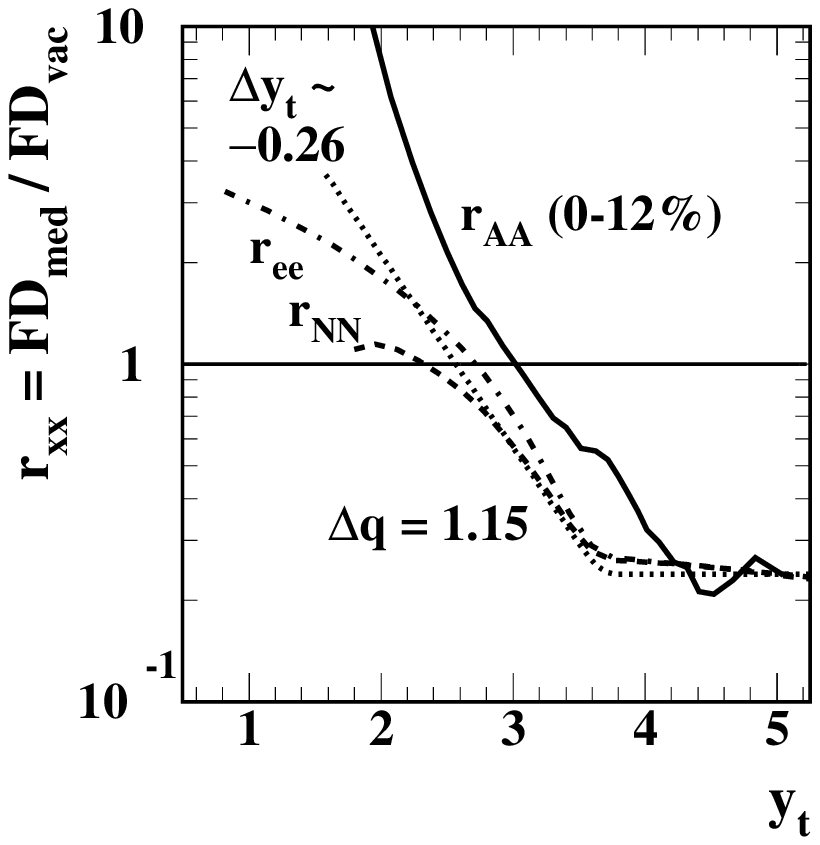}
\includegraphics[width=1.65in,height=1.75in]{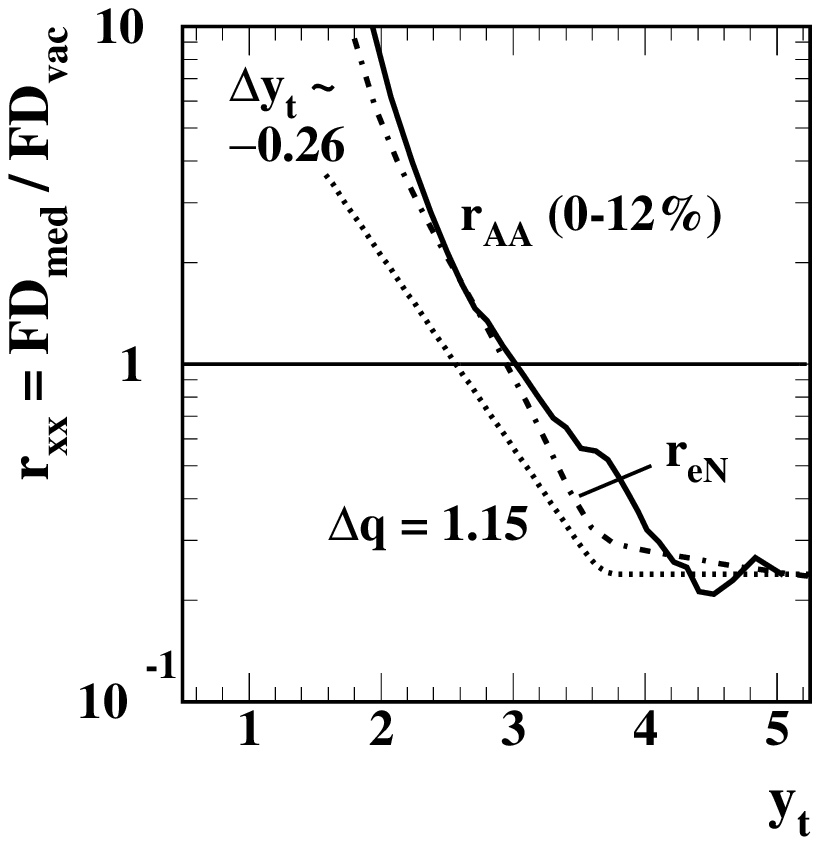}
\caption{\label{raas} 
Left: Calculated hard-component ratios $r_{xx}$ compared to measured $r_{AA}$ for 0-12\% central 200 GeV Au-Au collisions (bold solid curve).
Right: Same as the left panel except for newly-defined ratio $r_{eN} = H_\text{ee-med} / H_\text{NN-vac}$ which compares well with the Au-Au data.
} 
\end{figure}

Figure~\ref{raas} (left panel) shows calculated FD ratios $r_{NN}$ (dashed curve, p-p FFs) and $r_{ee}$ (dash-dotted curve, e-e FFs). The solid curve is the measured $r_{AA}$ from central (0-12\%) Au-Au collisions at 200 GeV~\cite{2comp}. $\Delta q \sim 1.15$ for $H_{ee-med}$ and $H_{NN-med}$ was adjusted to obtain the correct large-$y_t$ suppression for that centrality. The reference for $r_{AA}$ is hard-component model function $H_{GG}$. The dotted curve is the ratio reference obtained by shifting $H_{GG}$ on $y_t$ by $\Delta y_t \sim -0.26$ (negative boost). As noted in~\cite{2comp} the simple negative-boost model does not describe the Au-Au data. But the e-e and N-N ratios also do not describe the data.

Figure~\ref{raas} (right panel) introduces a novel concept. Instead of comparing the calculated in-medium FD for N-N (averaged within A-A collisions) with the in-vacuum FD for N-N or similarly comparing e-e with e-e as in the left panel, the in-medium FD for e-e is compared with the  in-vacuum FD for N-N by defining ratio
\bea
r_{eN} &=& \frac{H_\text{ee-med}}{H_\text{NN-vac}}.
\eea
Calculated $r_{eN}$ describes the measured $r_{AA}$ well over the entire fragment momentum range. The question then arises how to interpret the result.

\subsection{Revised FD reference for nuclear collisions} \label{revised}

Fig.~\ref{raas} (right panel) implies that the FD reference for nuclear collisions should be reconsidered. The hard component for central Au-Au collisions appears to be well-described by $H_\text{ee-med}$ with the BW ``energy loss'' mechanism applied to e-e FFs.
The p-p hard component deviates strongly from $H_\text{ee-vac}$ in Fig.~\ref{eefd} (right panel) and may be strongly suppressed at smaller $y_t$. The combination suggests that 
p-p hard component  $H_\text{NN}$ is not the correct reference for A-A collisions as assumed implicitly in defining conventional ratio $R_{AA}$. The proper {in-vacuum} reference for nuclear collisions is actually $H_\text{ee-vac}$. 

\begin{figure}[h]
\includegraphics[width=.47\textwidth]{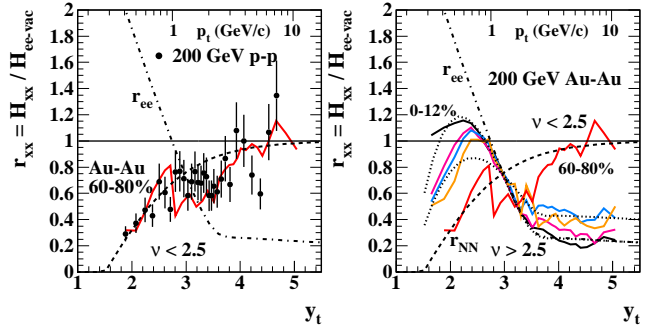}
\caption{\label{rnew}  (Color online)
Left: Hard-component ratios $r_{xx}$ based on the calculated reference $H_\text{ee-vac}$ determined with in-vacuum $e^+$-$e^-$  FFs. The data are for $H_\text{pp}$ from p-p collisions (solid points) and $H_\text{AA}$ from Au-Au collisions with $\nu < 2.5$ (solid curve).  $r_{ee}$ includes the calculated $H_\text{ee-med}$ from in-medium $e^+$-$e^-$  FFs (dash-dotted curve). $r_{NN}$ includes the calculated $H_\text{NN-vac}$ from in-vacuum p-\=p  FFs (dashed curve).
Right: Same as left panel but for $H_\text{AA}$ from Au-Au collisions with $\nu > 2.5$. The dotted curves are discussed in the text.
} 
\end{figure}

Figure~\ref{rnew} (left panel) shows FD ratios redefined in terms of the ee-vac reference: $H_\text{pp}$ (p-p data -- points),  $H_\text{AA}$ (peripheral Au-Au data -- solid curve) and calculated $H_\text{ee-med}$ (dash-dotted curve) and $H_\text{NN-vac}$ (dashed curve) all divided by reference $H_\text{ee-vac}$. The strong suppression of p-p and peripheral Au-Au data apparent at smaller $y_t$ results from the cutoff of p-p FFs noted above. 

Figure~\ref{rnew} (right panel) shows measured $H_\text{AA}/H_\text{ee-vac}$ for more-central Au-Au collisions (solid curves) above a transition point on centrality at $\nu \sim 2.5$. Centrality measure $\nu \equiv 2\, n_{bin} / n_{part}$ is the mean participant-nucleon path length in number of N-N collisions. For the Au-Au collisions in Fig.~\ref{rnew} $\nu$ values for the five centralities are 1.93, 2.83, 3.92, 4.87, 5.5, where $\nu \sim 1.25$ is N-N collisions and $\nu \sim 6$ is $b = 0$ Au-Au collisions~\cite{centmeth,2comp}. From $\nu$ = 1.98 to $\nu$ = 2.83 there is a dramatic change in the hard component. At the transition point $\nu \sim 2.5$ $n_{part} = 40$ (out of 382) and $n_{bin} = 50$ (out of 1136)~\cite{centmeth}.

\begin{table}[h]
\caption{\label{tab1} Parameters for $H_\text{xx}$ {\em vs} centrality}
\begin{center}
\begin{tabular}{|c|c|c|c|c|c|} \hline
 centrality & $\Delta q$ & $y_0 = \xi_y$ & $y_{cut}$ & $E_{cut}$ (GeV) & xx  \\ \hline\hline
reference & 0.0 & 0.0 & 3.75 & 3.0 & ee-vac  \\ \hline 
 1 & 0.0 & 1.5 & 3.75 & 3.0 & NN-vac  \\ \hline
 2 & 0.7 & 1.25 & 3.7 & 2.85 & $\cdots$  \\ \hline
 5 & 1.15 & 1.25 & 3.65 & 2.7 & $\cdots$ \\ \hline
reference & 1.15 & 0.0 & 3.65 & 2.7 &  ee-med  \\ \hline
 \end{tabular}
\end{center}
\end{table}

Table~\ref{tab1} shows parameters for calculated FDs $H_\text{xx}$ {\em vs} Au-Au centrality plotted as ratios to common reference $H_\text{ee-vac}$ in Fig.~\ref{rnew} (right panel). $\Delta q$ is determined in all cases by the region above $y_t = 4$ (conventional ``suppression''). $y_{cut}$ or $E_{cut}$ is determined by the slope in the intermediate region near $y_t = 3$. FF cutoff parameters $y_0 = \xi_y$ are determined by structure to the left of $y_t = 2.5$. 
The first line describes reference $H_\text{ee-vac}$ ($r_{xx} = 1$) with cutoff energy $E_{cut} = 3$ GeV. The second line describes $H_\text{NN-vac}$ (dashed curve). The last line describes the limiting case of $H_\text{ee-med}$ for central Au-Au collisions (dash-dotted curve) with $E_{cut} = 2.7$ determined by the data near $y_t = 3$. The lighter dash-dotted curve is the same with $E_{cut} = 3$ GeV for comparison. Dotted curves 2 and 5 are hybrid versions $H_\text{xx-med}$ with $y_0 = \xi_y$ adjusted to accommodate the data to the left of $y_t = 2.5$. 

The results can be interpreted as follows. With increasing centrality the splitting cascade is modified ($\Delta q$ increases from zero), suppression of FFs at smaller $y$ is reduced ($y_0 = \xi_y$ move toward zero) and parton spectrum cutoff $y_{cut}$ ($E_{cut}$) is also reduced, increasing $\sigma_{dijet}$ and the total minijet yield by 50\% as in Fig.~\ref{partspec} (left panel).  
Instead of invoking $r_{xx}$ ratios the measured spectrum hard components can be compared directly with calculated FDs to reveal fragmentation evolution.

\subsection{Fragmentation evolution}

\begin{figure}[h]
\includegraphics[width=.47\textwidth]{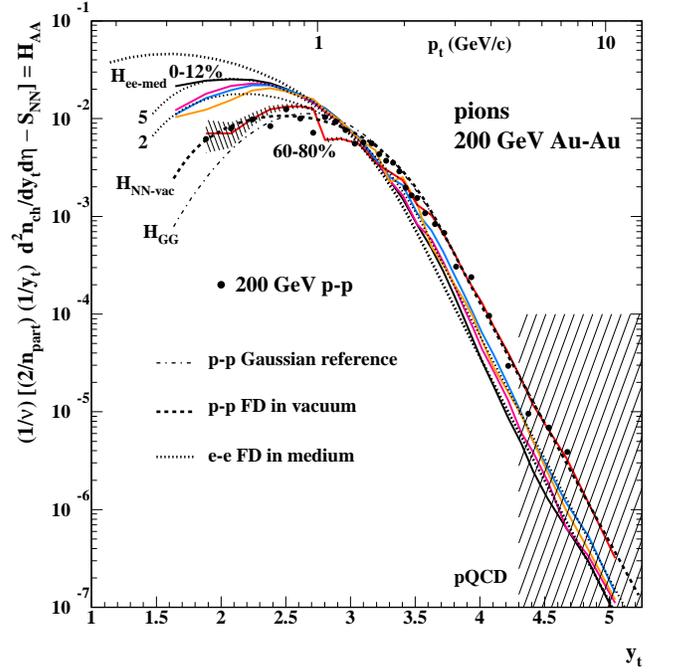}
\caption{\label{aafds}
(Color online) Measured spectrum hard components $H_{AA}$ for five centralities from 200 GeV Au-Au collisions (bold curves of several colors) and 200 GeV NSD p-p collisions (solid points) compared to calculated FDs for several conditions (vacuum, medium, $e^+$-$e^-$, p-\=p). The hatched region at upper left esimates the uncertainty due to the $S_{NN}$ subtraction common to all centralities. The hatched region at lower right denotes the interval conventionally allotted to pQCD.
} 
\end{figure}

Figure~\ref{aafds} shows spectrum hard components $H_{AA}$ (solid curves) for five centralities from 200 GeV Au-Au collisions~\cite{2comp}. This format shows $H_\text{AA}$ and related FD curves, whereas Fig.~\ref{aaspec} (left panel) shows $\nu\, H_\text{AA}$ including participant path length $\nu$. Since $\nu\, n_{part} / 2 = n_{binary}$ the hard components of un-normalized $y_t$ spectra (2D densities) scale proportional to $n_{binary}$ as expected for parton scattering and fragmentation in A-A collisions.

The points are hard-component data from 200 GeV NSD p-p collisions~\cite{ppprd}. The dash-dotted curve is the standard Gaussian+tail model function $H_\text{GG}$. FDs from the previous section are also shown. The dashed curve is $H_\text{NN-vac}$, and the upper (bold) dotted curve is $H_\text{ee-med}$ with $\Delta q = 1.15$, which nominally corresponds to the most-central Au-Au curve (0-12\%). The parton spectrum cutoff for $H_\text{ee-med}$ has been reduced from 3 GeV ($y_{max} = 3.75$) to 2.7 GeV ($y_{max} = 3.65$) to match the central Au-Au hard component near $y_t = 3$. The two thinner dotted curves labeled 2 and 5 (Au-Au centralities) are $H_\text{ee-med}$ with cutoff parameters $y_0 = \xi_y$ reduced to accommodate the data below $y_t = 2.5$.

The p-p and peripheral Au-Au data are consistent with $H_\text{NN-vac}$ by construction. Above a transition point on centrality ($\nu \sim 2.5$) $H_{AA}$ transitions from  $H_\text{NN-vac}$ toward $H_\text{ee-med}$ over most of the $y_t$ range, with residual deviations confined to smaller $y_t$. For more-central collisions agreement of $H_\text{AA}$ with the limiting $H_\text{ee-med}$ curve extends toward the limits of accepted $y_t$.


\subsection{Restoration of the FF base in A-A collisions}

Below centrality $\nu = 2.5$ we observe strong fragment suppression at smaller $y_t$ relative to the $H_\text{ee-vac}$ reference. Above that point there is asymptotic approach to $H_\text{ee-med}$ (bold dotted curve in Fig.~\ref{aafds}), with strong enhancement at smaller $y_t$ and the expected suppression at larger $y_t$ characterized as ``jet quenching''. The FD modification in central Au-Au collisions is approximately consistent with the BW in-medium modification of  e-e FFs~\cite{b-w}. Those trends suggest the following scenario: 

\begin{itemize}  

\item The bases of fragmentation functions (jets) in p-p and peripheral A-A collisions are missing from reconstructed jets and the spectrum hard component compared to in-vacuum e-e FFs

\item Above a transition point on A-A centrality fragmentation changes dramatically

\item The bases of A-A FFs are partially restored to compatibility with e-e FFs

\item A-A FFs are modified in a manner compatible with a pQCD description of FF medium modification

\item Most of the underlying parton spectrum does not change with Au-Au centrality; no scattered partons are lost to the final state (thermalized)

\item However, the low-energy cutoff---3 GeV for N-N collisions---falls to 2.7 GeV for central Au-Au collisions, increasing the minijet cross section by 50\%

\end{itemize}

The transition of the Au-Au spectrum hard component near $\nu \sim 2.5$ corresponds to a similar sharp transition observed in minijet angular correlations at 200 GeV~\cite{daugherity}, again consistent with equivalence of the spectrum hard component and minimum-bias parton fragmentation.

\section{Related aspects of A-A collisions}

Given a better understanding of scattered-parton spectra and fragmentation we consider the consequences in A-A collisions for parton distributions in the initial state and hadron distributions in the final state.

\subsection{pQCD parton spectra and A-A initial conditions} \label{disctheory}

Parton spectra determined {\em ab initio} from pQCD are used to estimate the initial conditions for heavy ion collisions at RHIC. In~\cite{cooper}  a jet spectrum was obtained from a parton spectrum defined in terms of a pQCD parton differential cross section and nucleon structure functions
\bea \label{cooper}
\frac{dn_{jet}}{dp_t}&=& K\, T(0)\,\frac{d\sigma_{dijet}}{dp_t},
\eea
where $K = 2$ (``higher-order contributions'') and $T(0)$ is the ``nuclear geometrical factor'' $= 9A^2 /8 \pi\, R_A^2$, with $R_A = 1.1\, A^{1/3}$ and $T(0) = 34$ mb$^{-1}$ for central Au-Au collisions. The pQCD spectrum was determined as in this analysis for central Au-Au collisions at 200 GeV where  $n_{binary} = 1136$ and $\sigma_{NSD} = 36.5$ mb. $T(0)$ can then be compared to $n_{binary} / \sigma_{NSD}$ = 31 mb$^{-1}$. Extracting $dn_{jet} / dp_t$ data from~\cite{cooper} (Fig.~1, 2-10 GeV) and rearranging Eq.~(\ref{cooper}) to
\bea
\frac{d\sigma_{dijet}}{dy_{max}}   &=& \frac{p_t}{K\, T(0)}\,\frac{dn_{jet}}{dp_t}
\eea
[with $y_{max} = \ln(2p_t / m_\pi)$] we recover the pQCD parton spectrum, plotted as the bold dotted curve in Fig.~\ref{partspec} (left panel),  corresponding closely (near the peaks) to the spectra inferred from FD data.  

Whereas FD data and this analysis imply that the parton spectrum terminates near 3 GeV with a 2.5-4 mb total cross section, the spectrum in~\cite{cooper} was integrated down to 1 GeV to estimate a parton (minijet) density $dn_{jet} / dy_z = 750$ for central Au-Au collisions. The lower cutoff was justified by saturation-scale arguments~\cite{cgc} (cf. App.~\ref{cgc}). Given $4\pi$ rapidity interval = 7, $n_{binary} \sim 1136$  and two jets per parton collision the implied dijet total cross section is about 85 mb (and cf. Fig.~\ref{partspec} -- left panel), larger than the {\em total} N-N cross section and more than $20$ times the value we infer from hadron spectrum data. Based on a 4 mb jet cross section for central Au-Au collisions we expect $35\pm 9$ minijets in one unit of $\eta$, consistent with minijet correlation analysis~\cite{jeffpp,daugherity}.

Arguing by analogy, a parton spectrum observed via charged hadrons should be terminated by the available density of hadronic final states, the same mechanism that terminates a splitting cascade in jet formation. An isolated parton scatter (e.g. in 200 GeV p-p collisions where at most one parton scatter occurs) should not proceed unless there is at least one hadronic final state available. 

It could be argued that in more-central Au-Au collisions the parton spectrum is substantially altered (cutoff extended to much lower energies) by the environment, that partons scatter and rescatter until hadrons finally emerge from a collective medium (cf. Sec.~\ref{cgc}). But that is not what we observe in direct comparisons of calculated FDs with measured spectrum hard components and correlations. We do see a modest decrease (10\%) in the cutoff energy with corresponding 50\% increase in the minijet cross section, as in Fig.~\ref{partspec} (left panel, dash-dotted curve). However, strong constraints from the hadron density of states apparently still apply to individual parton scatters. Rescattering of partons and hadrons or ``constituent quark recombination'' from a medium are contradicted by hard-component data. 


\subsection{Charged-hadron and total-$p_t$ production} \label{production}

Models for initial-state parton scattering and fragmentation should confront measured features of the final state. Beyond spectrum hard components this analysis can be compared with soft and hard components of integrated charged-hadron yields and total $p_t$. The two-component model for A-A particle production (with impact parameter $b$ and participant path-length $\nu$) is 
\bea
\frac{2}{n_{part}}\frac{dn_{ch}}{d\eta} &=& \frac{dn_{s}}{d\eta} + \nu \frac{dn_{h}}{d\eta} \\ \nonumber
&=& 2.5 + \frac{\nu}{\sigma_{NSD}} \frac{\sigma_{dijet}(b)}{\Delta\eta}\, \epsilon_{xx}(b)\,\bar n_{dijet}(b) \\ \nonumber
&=& 2.5 + 0.02 ~\text{   p-p collisions}
\\ \nonumber
&\approx& 2.5 + 1.1 ~~b = 0 \text{   Au-Au collisions},
\eea
defining $dn_h/d\eta$ (integral of $H_{AA}$ on $y_t$) as the average hard component for a single N-N collision within an A-A collision. The soft-component multiplicity retains the \mbox{p-p} value 2.5.
The jet cross section (and therefore minijet number) increases by about 50\% with A-A centrality due to reduction of the parton spectrum cutoff, and the mean dijet multiplicity increases about 3-fold. In those A-A collisions with multiple jet pairs $\epsilon_{AA} \sim 2\, \epsilon_{pp}$, effectively doubling the number of observed jets per N-N collision.

A similar expression for the total-$p_t$ density is
\bea
\frac{2}{n_{part}}\,P_t &=& \frac{dn_{s}}{d\eta} \langle p_t \rangle_s + \nu \frac{dn_{h}}{d\eta} \langle p_t \rangle_h(b) \\ \nonumber
&=& 0.88 \text{  GeV/c} + 0.02 \text{  GeV/c} ~\text{   p-p} \\ \nonumber
&\approx& 0.88 \text{  GeV/c} + 0.5 \text{  GeV/c}~~~b = 0 \text{ Au-Au},
\eea
where $P_t$ is the total $p_t$ in one unit of $\eta$, $\langle p_t \rangle_s \sim 0.35$ GeV/c is fixed, and $\langle p_t \rangle_h(b)$ is $\sim 1$ GeV/c for p-p collisions but {\em decreases} with increasing A-A centrality to about $0.5$ GeV/c due to medium modification of the FFs. Note that 1.38 GeV/c / 3.6 = 0.38 GeV/c, consistent with the $\langle p_t \rangle$ centrality variation for Au-Au collisions~\cite{2comp}:  $\langle p_t \rangle$ increases with A-A centrality through a maximum, and then decreases due to the FF medium modification.

In contrast, the hard-component contribution from~\cite{cooper} predicting $dn_{jet}/dy = 750$ (parton spectrum cutoff at 1 GeV) for central Au-Au collisions at 200 GeV should be
\bea
\frac{2}{n_{part}} \,P_t &=&\frac{2}{3} \times \frac{750}{191} \times 1 \text{ GeV/c} \\ \nonumber
&\sim& 2.5 \text{ GeV/c} ~~b = 0 \text{ Au-Au collisions},
\eea
exceeding by five times the hard-component $P_t$ observed in data. Measured charged-hadron minijets corresponding to a 4 mb jet cross section account for all  hard-component $p_t$ production. That reasoning closes the loop among correlations, single-particle spectra and pQCD.


\section{Systematic uncertainties} \label{errors}

This analysis compares calculated FDs with measured spectrum hard components over a large kinematic range. The critical elements of the comparison are the parton spectrum, parametrized FFs from e-e and p-p collisions, an ``energy loss'' model applied to FFs, and spectrum hard components from p-p and Au-Au collisions. 

\subsection{Parton spectrum}


In this analysis a jet cross section is obtained by comparing a pQCD calculation with a minimum-bias fragment distribution rather than with a reconstructed-jet spectrum. The jet differential cross section on $\eta$ and its relative error can be extracted from Eq.~\ref{jetcross}
\bea
 \frac{\sigma_{dijet}}{\Delta \eta} \approx \frac{d\sigma_{dijet}}{d\eta} &\equiv& \frac{\sigma_{NSD}}{\epsilon(\delta \eta,\Delta \eta)\, \bar n_{dijet}}\frac{dn_h}{d\eta}.
\eea
The parameter values with estimated systematic errors are $\sigma_{NSD} = 36.5\pm 2.5$ mb~\cite{nsd}, $dn_h/ d\eta = 0.02 \pm 0.0025$ (\cite{ppprd} and this analysis), $\bar n_{dijet} = 3.0 \pm 0.5$ (CDF FFs~\cite{cdf2} and systematic studies for this analysis) and $\epsilon(\delta \eta=1) = (0.8 \pm 0.1)\{1 + 0.1 \pm 0.025\}/2$ (estimate based on the angular widths of observed minijets and the small curvature of the away-side ridge within $|\eta| < 1$~\cite{daugherity}). The combined error for the differential cross section is then about 25\%.



Given the basic power-law form with cutoff, the parton spectrum is defined by three parameters---$y_{cut}$ ($E_{cut}$), $n_{QCD}$ and amplitude $A_{y_{max}}$. The three are simply related to $\sigma_{dijet}$ in Sec.~\ref{parspec}. $n_{QCD}$ and amplitude $A_{y_{max}}$ are determined by comparison to the p-p hard component up to 7 GeV/c, as in Fig.~\ref{eloss3} (right panel), and the peripheral Au-Au hard component up to 10 GeV/c, as in Fig.~\ref{aafds}. Values $n_{QCD} = 7.5\pm 0.5$, $A_{y_{max}} = (8\pm1) \times 10^{9}$ and $y_{cut} = 3.75 \pm 0.05$ ($E_{cut} = 3.0\pm 0.15$ GeV) are consistent with $\sigma_{dijet} = 2.5\pm 0.6$ mb.  

While the spectrum cutoff for p-p collisions is well-defined by data the effective cutoff for e-e FFs is less well-defined. The cutoff used for the in-vacuum e-e reference FD is by definition the same as inferred from the p-p hard component. The cutoff used for centralities above the transition in Au-Au collisions was reduced to 3.65 ($E_{cut} \sim 2.7$ GeV) to match the corresponding FDs to Au-Au hard components in the region near $y_t \sim 3$ ($p_t = 1.5$ GeV/c). The uncertainty in the modified cutoff energy determined by central Au-Au data is 0.15 GeV [cf. Fig.~\ref{rnew} (right panel) thick and thin dash-dotted curves].

\subsection{Fragmentation functions}

The beta parametrizations in~\cite{ffprd} describe $e^+$-$e^-$ FF data to their error limits from 10 to 200 GeV energy scale (dijet energy) and for all fragment momenta from 0.1 GeV/c to the parton momentum. Measured absolute fragment yields (dijet multiplicities) are also well described. The parametrization allows reliable extrapolation down to $Q = 4$-6 GeV ($E_{jet} = 2$-3 GeV).

The p-\=p FF parametrization used in this analysis is less-well developed. Measured FFs plotted on rapidity for the first time in~\cite{ffprd} revealed substantial systematic deviations at small fragment momenta (suppression) relative to $e^+$-$e^-$ FFs, although both are reported to be well-described by the MLLA~\cite{cdf1,opal}. This analysis suggests that the deviations are due to physical differences, not detection or reconstruction inefficiencies.

Proper description of the suppression of p-\=p FFs is essential for the comparison between calculated FDs and spectrum hard components. E.g., the suppression at $Q = 6$ GeV amounts to more than a factor 2 decrease in dijet multiplicity. Yet the p-\=p FFs so described lead to good agreement with theoretical jet cross sections.

A two-parameter $\tanh$ cutoff function with parameters $(y_0,\xi_y)$ applied to the $e^+$-$e^-$ FF parametrization describes CDF FFs well, especially for smaller dijet energies. The FF cutoff influences the shape of calculated FDs near the spectrum mode, as does the parton spectrum cutoff $y_{cut}$. The parameters could interact, for example in describing the p-p hard component. In Fig.~\ref{ppffs2} (right panel) the dotted curve shows the result of reducing parameter $y_0$ from 1.6 to 1.4 (which would strongly disagree with measured p-\=p FFs). The difference can be compared with the result of changing $y_{cut}$ by 0.1 in Fig.~\ref{pspec} (left panel). The shape changes are easily distinguished at the few-percent level.

\subsection{Energy-loss model}

A simple model of FF medium modification was adopted for this study. It has the advantage that alteration of one parameter ($q$) in the beta-distribution description of $e^+$-$e^-$ FFs reproduces the BW pQCD model. An accurate description of ``medium modification'' over the complete range of fragment momenta is thus possible. 


``Energy-loss'' parameter $\Delta q$ determined by spectrum suppression at larger $p_t$ produces large effects at smaller $p_t$ which were unanticipated but describe Au-Au data well. $\Delta q$ modifies beta distribution control parameters $(p,q)$ derived for $e^+$-$e^-$ FFs~\cite{ffprd}. A value  $\Delta q = 1.15 \pm 0.2$ for central Au-Au collisions describes suppression of the hard component up to 10 GeV/c  to the error limits of data. The resulting e-e FFs for 14 and 200 GeV agree with the BW result to a few percent over the range of validity of the MLLA FFs used in the BW study.

\subsection{Spectrum hard components}

Spectrum hard components are obtained as differences between measured spectra and soft component $S_{xx}$ inferred from the multiplicity (in p-p collisions~\cite{ppprd}) or centrality (in Au-Au collisions~\cite{2comp}) variation of a spectrum ensemble. Systematic uncertainties derive from the measure spectra and from the subtracted soft component.

The limit procedure used to obtain the soft component (a form of Taylor expansion) suppresses systematic uncertainties derived from data and the inferred structure of $S_{xx}$ in the difference defined as the hard component. What error remains is common to all extracted hard components. Differential centrality variation of hard components is then relatively free of systematic error because of the double suppression of common-mode error. The absolute uncertainty due to the $S_{xx}$ subtraction, common to all centralities, is estimated by the upper-left hatched region in Fig.~\ref{aafds}.

\section{Discussion}

This analysis provides direct comparison of the complete hadron-spectrum hard component from nuclear collisions with parton fragment distributions derived per pQCD from fragmentation functions and a common parton spectrum. Evolution of the spectrum hard component with Au-Au centrality reveals that the FF ensemble undergoes a transition from in-vacuum p-\=p FFs to in-medium $e^+$-$e^-$ FFs at a specific centrality. Details and implications of the analysis are now discussed.

\subsection{Hard component as fragment distribution}


Spectrum hard components previously extracted from p-p and Au-Au collisions are identified as minimum-bias fragment distributions or FDs. Calculated FDs are in turn  generated with improved accuracy by folding beta distribution FF parametrizations with a power-law parton spectrum model. Comparison with the \mbox{p-p} hard component determines a spectrum cutoff and minimum-bias jet cross section consistent with pQCD and theoretical analyses of UA1 $E_t$ clusters. 

Such comparisons confirm in detail that a pQCD description of parton scattering and fragmentation is valid down to at least 3 GeV parton energy, the {\em effective} lower limit for partons fragmenting to charged hadrons. Further confirmation comes from correlation hard components which reveal unambiguous jet angular correlations down to small hadron $p_t$ ($\sim$ 0.1-0.3 GeV/c).

These combinatoric methods have several advantages compared to the UA1 event-wise $E_t$ analysis. The UA1 analysis imposed a jet definition (cone jet finder) with attendant bias. A substantial background came from the underlying event (1-2 GeV within the jet cone). In contrast, the hard component of particle spectra reveals the lower edge of the FD with relatively small bias (the subtracted soft component is defined by a limiting procedure unrelated to hard-component interpretation). The parton spectrum cutoff is thus accurately determined. 

\subsection{New fragmentation phenomenology}

We observe that p-p FFs are systematically different from e-e FFs (e.g., OPAL/TASSO vs CDF), but the difference is apparent only when FF data are plotted on rapidity $y$ down to small fragment momenta as in~\cite{ffprd}. The hard component of 200 GeV p-p collisions is well described by folding parametrized p-\=p FFs with a power-law parton spectrum terminating near 3 GeV. Since the p-p {\em minimum-bias} hard component is well described by p-\=p FFs jet-reconstruction and particle-detection inefficiencies seem not to cause the differences. 

We therefore conclude that p-p FFs are indeed suppressed at smaller fragment momenta relative to e-e FFs, implying that novel QCD physics emerges in p-p collisions, and the p-p hard component is not the proper FD reference for nuclear collisions. Use of the p-p FD (or worse, the entire p-p $p_t$ spectrum) as a reference appears to be misleading. The correct FD reference for all nuclear collisions is obtained by folding in-vacuum e-e FFs with a power-law parton spectrum. A unique aspect of the present analysis (compared to conventional NLO FD calculations which follow the same approach) is the accuracy and extent of the e-e FF parametrization in~\cite{ffprd}.

The centrality evolution of FDs is studied in this analysis directly with spectrum hard components and indirectly with ratio measures. A new phenomenon has emerged:  At a particular centrality there is a sharp transition from the suppressed p-p FD to a medium-modified e-e FD with at least three-fold increase in dijet multiplicity. The p-p FD observed in peripheral Au-Au becomes in central Au-Au not an ``energy-loss'' p-p FD but a ``medium-modified'' e-e FD, and the effective cutoff energy of the parton spectrum is reduced by about 10\%, increasing the jet total cross section by about 50\%. 

\subsection{Physical interpretation -- p-p}

The hard component from 200 GeV p-p collisions is well described by a parton spectrum folded with p-\=p fragmentation functions measured at FNAL. The FF suppression at smaller $p_t/y_t$ appears to be a real loss of part of the jet (the base), albeit a small fraction of the parton energy. 
Where is the missing jet base?

For dijets in e-e collisions the color field is localized and continuous along the $q$-$\bar q$ axis. The fragment density on $y_z$ (from momentum component $p_z$) along the color-dipole axis is uniform near the parton CM~\cite{aleph}. Corresponding FFs on $y$ (from {\em total} momentum $p$) as in Fig.~\ref{eeffs} (left panel) fall smoothly to zero at $y_{min}$ ($p_t \sim 0.05$ GeV/c). In p-p collisions the ``bases'' of p-p FFs are missing relative to e-e FFs, implying that the fragment distribution on $y_z$ acquires a ``hole'' at the parton CM. The hole in the dipole distribution suggests that the scattered-parton pair is not color connected, leading to these observations:

1) Parton ``scattering'' in p-p collisions must proceed by exchange of a color singlet between nucleons (a variant of Pomeron exchange). All momentum transfers between nucleons are colorless, consistent with no color connection between scattered partons. 

2) A scattered parton remains color connected to the parent nucleon.  Part of the color field deviates from the parton-parton axis (follows the projectile nucleon). 

3) Fragmentation responds to the color connection, and some fragments are shifted away from the parton-parton axis (away from p-p mid-rapidity). FF bases are then missing from observed jets and hard component.

The color-connection mechanism may explain suppression of the bases of  p-p FFs. If so, the larger the $x$ of the scattered parton the larger should be the suppression or loss from the FF, as suggested by CDF FFs~\cite{cdf1,cdf2}.

\subsection{Physical interpretation -- Au-Au}

In more-central Au-Au collisions (above the sharp transition) the FD approaches a limiting case corresponding to ``medium-modified'' e-e FFs. Instead of suppression at smaller $y_t$ there is enhancement. Fragmentation is modified, and jet bases are partially restored. The suppression mechanism proposed in the previous subsection implies that color connection in more-central Au-Au collisions transitions from the N-N case toward the color dipole of $e^+$-$e^-$ collisions. However, the splitting cascade is also modified within the Au-Au collision context.

Minijet angular correlations reveal that restored jet bases are strongly elongated on $\eta$ relative to the larger-$y_t$ part of the FF~\cite{axialci,daugherity}. Triggered jet analysis refers to FF components as ``jet'' and ``ridge''~\cite{putschke}. From this analysis the $\eta$ elongation can be explained by residual color connection to parent nucleons and resulting distortion of fragment distributions relative to the back-to-back dijet configuration in $e^+$-$e^-$ collisions. The larger the $x$ of the struck parton the stronger should be the $\eta$ elongation.



The underlying parton spectrum remains the same except for the low-energy cutoff, which drops from 3 GeV to 2.7 GeV, increasing the jet cross section by 50\%. Combined with a 3-fold increase in minijet multiplicity and doubling of dijet efficiency $\epsilon$ the total fragment multiplicity per unit $\eta$ per N-N collision increases by a factor 9 in central Au-Au compared to p-p collisions. The spectrum cutoff may be reduced in more-central Au-Au collisions because the fragmentation process is altered: the hadron density of states becomes larger and the effective cutoff can then move down. There is no suggestion of parton loss by absorption or multiple scattering in a medium.

\subsection{Implications for conventional energy-loss models}

In a summary of high-$p_t$ physics at RHIC and LHC~\cite{est} it was reported that
``observation of the strong suppression of high $p_t$ hadrons'' motivates the choice of `hard probes'' (energetic partons) to ``characterize the deconfined medium'' at RHIC. Energetic partons ``are expected to lose energy through collisional energy loss and medium-induced gluon radiation...known as jet quenching.'' Effects of a flowing medium may also be revealed~\cite{flow-wind}. But energy-loss schemes based on random multiple scattering of a leading parton are falsified by minijet angular-correlation data, especially {\em reduction} of the minijet azimuth width with increasing A-A centrality and survival of essentially all scattered partons above 3 GeV as jet correlations in the hadronic final state. 

Earlier high-$p_t$ results from RHIC such as ``disappearance of the away-side jet'' from dihadron correlations~\cite{disappear} and ``jet quenching'' from spectra~\cite{starraa} were interpreted to imply absorption of most jets (including minijets) in central collisions within an ``opaque core,'' observed jets being restricted to production in a surface layer. Those conclusions result in part from limitations of the measures employed. From this analysis we find that all jets survive to particle detection, albeit some are modified.

A measured negative shift of the {\em entire} FD on $y_t$ corresponds at most to a 25\% reduction of the {leading-parton} energy (for central Au-Au). Because the parton spectrum falls rapidly the shift does correspond to a 5-fold reduction of the FD at larger $p_t$. But energy is not necessarily lost from integrated FFs, and the total hadron multiplicity is not suppressed, instead it is shifted to a different part of the $p_t$ spectrum and actually increases.

Jet estimation by ``leading particle'' biases the jet structure (and imposes strong limitations on statistical power and accurate model tests), thus motivating full event-wise jet reconstruction. Regarding event-wise reconstruction ``...`jets' at RHIC are very complicated objects which make them {\em impossible to disentangle from the `background'}\, '' [emphasis added]. That may be true for event-wise reconstruction (except cf.~\cite{joern-jetff,mheinz}), but unbiased jet reconstruction is {\em in effect} accomplished on a combinatoric basis {\em via} hard components of nuclear spectra (FDs) and minijet correlations. The observed structure can be unfolded to reveal the effective parton spectrum and FF ensemble as modified in nuclear collisions.

``High-$p_t$ particle production in proton-proton collisions'' is said to ``provide the baseline `vacuum' reference to heavy-ion collisions to study the QCD medium properties''~\cite{est}. But in this analysis we find that mistaking the p-p hard component (or worse, the entire p-p spectrum) for an FD reference can distort ``medium'' effects at smaller $p_t$. Comparison of spectra to NLO FDs only above 2-3 GeV/c can also be misleading. The most significant modifications to fragmentation lie below that interval. Ironically, ``{\em soft} probes'' (minijets) reveal the most important details of QCD collision dynamics.

\subsection{Implications for hadrochemistry}

Strangeness and heavier flavors require exceptional energy densities for exceptional production (relative to a statistical-model reference). Exceptional energy densities are accessible within minijets. Systematic strangeness trends, described as "strangeness suppression" in p-p collisions~\cite{suppress} and "strangeness enhancement" in more-central A-A collisions~\cite{enhance},  suggest that minijet production may influence heavy-flavor (s, c, b) abundances in nuclear collisions. Because FF bases are suppressed in p-p collisions strangeness production is suppressed as well. In more-central A-A collisions the FF base is restored  (even enhanced), and strangeness (and heavier flavors) are enhanced as well. In that scenario strangeness production would closely follow minijet production and ``energy-loss'' trends. A possible correspondence of minijet systematics with heavy-flavor abundances should be carefully considered before invoking the context of a thermalized bulk medium, i.e., canonical suppression and enhancement within an equilibrated statistical ensemble.

\subsection{Implications for saturation-scale arguments} \label{cgc}

The  {\em saturation-scale} model (SSM) (e.g.~\cite{cgc}) is a limiting case of the two-component model of nuclear collisions in which the soft component disappears, all particle/$p_t/E_t$ production proceeds via the hard component, and all antecedent scattered partons are thermalized prior to hadronization. No predictions are made for hadronic correlations, and no justification is given for assumed disappearance of the soft component.

According to the SSM hypothesis the parton spectrum from A-A collisions is said to saturate at a particular momentum scale $p_0$  (saturation scale) depending on CM energy and A. Spectrum saturation is attributed to saturation of the initial parton density in the projectiles. Because of the parton spectrum structure almost all particle/$p_t/E_t$ production should then correspond to the saturation scale. The SSM parton spectrum cutoff for RHIC Au-Au collisions is estimated to be $\sim1$ GeV~\cite{cgc}. 

Semihard parton scatters from elementary hadronic collisions are isolated ($\ll$ one per collision), and p-p data imply that the density of hadronic final states constrains parton scattering near midrapidity. There is a direct QM coupling between parton scattering and hadronization in nuclear collisions: the latter constrains the former. 

In contrast, according to SSM any parton scatter that can happen kinematically will happen, independent of the density of hadronic final states, because there is no direct coupling between parton scattering and hadronization. Partons scatter to an intermediate QCD state (expanding QCD medium, possibly thermalized) from which hadrons later emerge in a collective process.  The only constraint on parton scattering is the saturation limit of the {\em initial parton flux density} from the projectiles. 

Based on that reasoning the SSM parton spectrum is extended from 3 GeV down to 1 GeV, as illustrated in Fig.~\ref{partspec} (left panel), implying a {\em large} increase in scattered partons compared to isolated p-p collisions (e.g., 85 mb dijet cross section in central Au-Au collisions {\em vs} 2.5 mb). The final-state hadron/$E_t/p_t$  production in A-A collisions is then attributed entirely to the hard component. The large soft-component production is incorrectly attributed to the SSM-invoked parton spectrum extension below 3 GeV to buttress the saturation-scale argument. 

The SSM hypothesis is contradicted by two-component analysis of correlations and spectra, particularly the centrality dependence (as in the present analysis). As demonstrated in Sec.~\ref{production} the soft component (projectile nucleon fragmentation) dominates hadron/$p_t/E_t$ production even in central Au-Au collisions. The complementary hard-component production matches the $\sim 3$ GeV parton cutoff inferred from hard-component spectra.  



\section{Summary}

In this analysis fragment distributions measured as $p_t$ spectrum hard components are used to resolve and define both parton spectra and FF ensembles over large kinematic intervals. Parton fragmentation and ``energy loss'' in nuclear collisions are studied. The initial conditions of nuclear collisions, particularly the scattered-parton energy distribution relevant to pQCD and possibly hydrodynamics, are estimated with reduced ambiguity.

Parametrized p-\=p and $e^+$-$e^-$ fragmentation functions are folded with a parton spectrum model to produce fragment distributions compared to hard components of hadron spectra. Comparison of a calculated p-\=p FD to a measured p-p hard component determines a reference parton spectrum for all nuclear collisions. The inferred spectrum agrees quantitatively with pQCD calculations.

A theoretical model of FF medium modification is implemented by simple alteration of FF parametrizations. In-medium  p-\=p and $e^+$-$e^-$ FFs are folded with the reference parton spectrum to produce modified FDs in turn compared to spectrum hard components for several centralities of 200 GeV Au-Au collisions. Below a transition centrality hard components for peripheral Au-Au and p-p collisions are well-described by the in-vacuum p-\=p FD. Above the transition centrality hard-component data are well-described by in-medium $e^+$-$e^-$ FDs over the entire $p_t$ spectrum [0.3,10] GeV/c. 

The implications of the analysis are as follows: 1) The underlying scattered-parton spectrum changes little from p-p to central Au-Au collisions; the exception is reduction of the 3 GeV cutoff energy by 10\%. All partons survive to final-state manifestations, are not ``thermalized.'' 2) Fragmentation functions for p-p collisions are substantially different from those for $e^+$-$e^-$ collisions---the low-momentum base is suppressed in the former, comprising 30-70\% of the expected fragment number but only a few percent of the parton energy. 3) The missing FF base suggests that scattered partons from p-p collisions are not color connected to each other, that a scattered parton remains color connected to the parent projectile nucleon.  4) In more-central Au-Au collisions there is a sharp transition to in-medium  $e^+$-$e^-$ FDs, suggesting that the color-connection topology changes substantially. 

5) Central Au-Au collisions appear to be nearly transparent to energetic partons and even to hadron fragments with $p_t \sim 0.3$ GeV/c which remain correlated with the parent parton, are not rescattered. 6) A measured large increase in Au-Au minijet angular correlations above the sharp transition on centrality corresponds to the large increase in mean dijet multiplicity for in-medium $e^+$-$e^-$ FFs compared to in-vacuum p-\=p FFs inferred from this analysis. 7) The most significant alteration of parton fragmentation in nuclear collisions occurs below $p_t = 2$ GeV/c. 8) A proposed extension of the scattered-parton $p_t$ spectrum down to 1 GeV in central Au-Au collisions motivated by saturation-scale arguments is contradicted by spectrum and correlation data.

Nuclear collisions have been described quantitatively in terms of perturbative QCD. Nonperturbative aspects already present in p-p collisions (fragmentation function modification and parton spectrum termination) evolve with heavy ion collision centrality in intuitively reasonable ways, leading to better understanding of QCD dynamics over an extended space-time volume.

This work was supported in part by the Office of Science of the U.S. DoE under grant DE-FG03-97ER41020



\begin{thebibliography}{99}

\bibitem{heinz} E. Schnedermann, J. Sollfrank and U. Heinz, 
Phys. Rev. C {\bf 48}, 2462 (1993).

 \bibitem{finns} P.~Huovinen and P.~V.~Ruuskanen,
  Ann.\ Rev.\ Nucl.\ Part.\ Sci.\  {\bf 56}, 163 (2006).

\bibitem{bwfit}   F.~Retiere and M.~A.~Lisa,
  Phys.\ Rev.\  C {\bf 70}, 044907 (2004).

\bibitem{rom} P.~Romatschke and U.~Romatschke,
  Phys.\ Rev.\ Lett.\  {\bf 99}, 172301 (2007).

\bibitem{highpt} P.~Jacobs and X.~N.~Wang,
  Prog.\ Part.\ Nucl.\ Phys.\  {\bf 54}, 443 (2005).

\bibitem{tomo}  W.~A.~Horowitz and M.~Gyulassy,
  Phys.\ Lett.\  B {\bf 666}, 320 (2008);.

\bibitem{tomo2} C.~A.~Salgado and U.~A.~Wiedemann,
  Phys.\ Rev.\ Lett.\  {\bf 89}, 092303 (2002).

\bibitem{2comp} T.~A.~Trainor,
  Int.\ J.\ Mod.\ Phys.\  E {\bf 17}, 1499 (2008), arXiv:0710.4504.

\bibitem{axialci}   J.~Adams {\it et al.}  (STAR Collaboration),
  Phys.\ Rev.\  C {\bf 73}, 064907 (2006).

\bibitem{daugherity}  M.~Daugherity  (STAR Collaboration),
  J.\ Phys.\ G {\bf 35}, 104090 (2008).

\bibitem{hijscale}  Q.~J.~Liu, D.~J.~Prindle and T.~A.~Trainor,
  Phys.\ Lett.\  B {\bf 632}, 197 (2006).
 
\bibitem{ptedep}  J.~Adams {\it et al.}  (STAR Collaboration),
  J.\ Phys.\ G {\bf 33}, 451 (2007).

\bibitem{ptscale}
J. Adams {\it et al.} (STAR Collaboration),
J. Phys. G: Nucl. Part. Phys. {\bf 32}, L37 (2006).

\bibitem{ppprd} J.~Adams {\it et al.}  (STAR Collaboration),
  Phys.\ Rev.\  D {\bf 74}, 032006 (2006).

\bibitem{ffprd}  T.~A.~Trainor and D.~T.~Kettler,
  Phys.\ Rev.\  D {\bf 74}, 034012 (2006).

\bibitem{cdf1}  K.~Goulianos  (CDF Collaboration), Proceedings of the ``QCD and high energy hadronic interactions,'' XXXII Rencontres de Moriond, Les Arces, France, March 22-29, 1997, 
FERMILAB-CONF-97-145-E. 

\bibitem{cdf2} D.~Acosta {\it et al.}  (CDF Collaboration),
Phys.\ Rev.\ D {\bf 68}, 012003 (2003).

\bibitem{kll}  K.~Kajantie, P.~V.~Landshoff and J.~Lindfors,
  Phys.\ Rev.\ Lett.\  {\bf 59}, 2527 (1987).

\bibitem{ua1} C.~Albajar {\it et al.}  (UA1 Collaboration),
  Nucl.\ Phys.\  B {\bf 309}, 405 (1988).

\bibitem{sarc} I.~Sarcevic, S.~D.~Ellis and P.~Carruthers,
  Phys.\ Rev.\  D {\bf 40}, 1446 (1989).

\bibitem{jeffpp}  R.~J.~Porter and T.~A.~Trainor  (STAR Collaboration),
  PoS C {\bf FRNC2006}, 004 (2006);
R.~J.~Porter and T.~A.~Trainor  (STAR Collaboration),
  J.\ Phys.\ Conf.\ Ser.\  {\bf 27}, 98 (2005),
hep-ph/0506172.

\bibitem{newflow}  T.~A.~Trainor and D.~T.~Kettler,
  Int.\ J.\ Mod.\ Phys.\  E {\bf 17}, 1219 (2008).

\bibitem{durand} L.~Durand and H.~Pi,
  Phys.\ Rev.\  D {\bf 40}, 1436 (1989).

\bibitem{blaizot}  J.~P.~Blaizot and A.~H.~Mueller,
  Nucl.\ Phys.\  B {\bf 289}, 847 (1987).

\bibitem{lowpqcd}  C.~Adloff {\it et al.}  (H1 Collaboration),
  Nucl.\ Phys.\  B {\bf 497}, 3 (1997).

\bibitem{nayak} G.~C.~Nayak, A.~Dumitru, L.~D.~McLerran and W.~Greiner,
  Nucl.\ Phys.\  A {\bf 687}, 457 (2001). 

\bibitem{minijet} X.-N. Wang,  Phys. Rev. D {\bf 46}, R1900 (1992); X.~N.~Wang and M.~Gyulassy,
  Phys.\ Rev.\  D {\bf 44}, 3501 (1991).

\bibitem{hijingpp} X.~N.~Wang and M.~Gyulassy,
  Phys.\ Rev.\  D {\bf 45}, 844 (1992).

\bibitem{b-w}   N.~Borghini and U.~A.~Wiedemann,
  hep-ph/0506218.

\bibitem{kkp} B.~A.~Kniehl, G.~Kramer and B.~P\"otter,
  Nucl.\ Phys.\ B {\bf 582}, 514 (2000).

\bibitem{tasso} W.~Braunschweig {\it et al.}  (TASSO Collaboration),
Z.\ Phys.\ C {\bf 47}, 187 (1990). 

\bibitem{opal} M.~Z. Akrawy {et al.}  (OPAL Collaboration)
  {Phys. Lett.} B, {\bf 247}, 617 (1990).

\bibitem{lphd} Ya.~I.~Azimov, Yu.~L.~Dokshitzer, V.~A.~Khoze, S.~I.~Troyan, Z. Phys. C {\bf 27}, 65 (1985),  Z. Phys. C {\bf 31}, 213 (1986).

\bibitem{cdfmult} D. Acosta {\em et al.} (CDF Collaboration), 
Phys. Rev. Lett. {\bf 94}, 171802 (2005).

\bibitem{cdf3} A.~Safonov (CDF Collaboration), Proceedings of the ``International Euroconference in Quantum Chromodynamics,'' Montpellier, France, July 7-13, 1999, CDF Note 5147,  October 14, 1999.

\bibitem{nlo}   D.~de Florian and W.~Vogelsang,
  Phys.\ Rev.\  D {\bf 71}, 114004 (2005).

\bibitem{phenixnlo}  A.~Adare {\it et al.}  (PHENIX Collaboration),
  Phys.\ Rev.\  D {\bf 76}, 051106 (2007).

\bibitem{ordering}  P.~Abreu {\it et al.}  (DELPHI Collaboration),
  Phys.\ Lett.\  B {\bf 407}, 174 (1997).


\bibitem{panch} G.~Pancheri and Y.~N.~Srivastava,
  Phys.\ Lett.\  B {\bf 182}, 199 (1986).

\bibitem{parteloss}  X.~N.~Wang and X.~F.~Guo,
  Nucl.\ Phys.\  A {\bf 696}, 788 (2001).

\bibitem{opal200} G.~Abbiendi {\it et al.}  (OPAL Collaboration),
  Eur.\ Phys.\ J.\  C {\bf 27}, 467 (2003).

\bibitem{centmeth}   T.\,A.~Trainor and D.\,J.~Prindle, 
 hep-ph/0411217.

 \bibitem{cooper} F.~Cooper, E.~Mottola and G.~C.~Nayak,
  Phys.\ Lett.\  B {\bf 555}, 181 (2003).

\bibitem{cgc} K.~J.~Eskola, K.~Kajantie, P.~V.~Ruuskanen and K.~Tuominen,
  Nucl.\ Phys.\  B {\bf 570}, 379 (2000).

\bibitem{nsd} G.~J.~Alner {\it et al.}  (UA5 Collaboration),
  Z.\ Phys.\  C {\bf 32}, 153 (1986).

\bibitem{aleph}  D.~Buskulic {\it et al.}  (ALEPH Collaboration),
  Z.\ Phys.\  C {\bf 55}, 209 (1992).

\bibitem{putschke}  J.~Putschke,
  J.\ Phys.\ G {\bf 34}, S679 (2007); nucl-ex/0701074.

\bibitem{est}  M.~Estienne,
  arXiv:0810.1698.

\bibitem{flow-wind} N.~Armesto, C.~A.~Salgado and U.~A.~Wiedemann,
  Phys.\ Rev.\ Lett.\  {\bf 93}, 242301 (2004);
 N.~Armesto, C.~A.~Salgado and U.~A.~Wiedemann,
  Phys.\ Rev.\  C {\bf 72}, 064910 (2005).

\bibitem{disappear} C.~Adler {\it et al.}  (STAR Collaboration),
  Phys.\ Rev.\ Lett.\  {\bf 90}, 082302 (2003).

\bibitem{starraa} C.~Adler {\it et al.}  (STAR Collaboration),
  Phys.\ Rev.\ Lett.\  {\bf 89}, 202301 (2002).

\bibitem{mheinz}  M.~Heinz  (STAR Collaboration),
  arXiv:0809.3769.

\bibitem{joern-jetff} J.~Putschke  (STAR Collaboration),
  arXiv:0809.1419.

\bibitem{suppress} H.~J.~Drescher, J.~Aichelin and K.~Werner,
  Phys.\ Rev.\  D {\bf 65}, 057501 (2002).

\bibitem{enhance} E.~Andersen {\it et al.}  (WA97 Collaboration),
  Phys.\ Lett.\  B {\bf 449}, 401 (1999).

\end{thebibliography}
\end{document}